\documentclass[letterpaper]{article}

\usepackage{spconf}

\usepackage[utf8]{inputenc}
\usepackage[T1]{fontenc}
\usepackage{booktabs}
\usepackage{amsfonts}
\usepackage{amsmath}
\usepackage{amssymb}
\usepackage{microtype}
\usepackage{xcolor}
\usepackage{graphicx}
\usepackage{adjustbox}
\usepackage{multirow}
\usepackage{array}
\usepackage{algorithm}
\usepackage{algpseudocode}
\usepackage[hidelinks]{hyperref}
\usepackage{url}
\usepackage{cleveref}
\usepackage{placeins}
\usepackage{eso-pic}

\title{LIFT-SE: Linguistic Inference Followed by Flow Transformation for Generative Speech Enhancement}

\name{Haoyin Yan \quad Chengwei Liu \quad Zheng Xue \quad Xiaotao Liang \quad Jifa Cai \quad Zeyu Zhao \quad Jingjing Wang}
\address{Qwen Business Unit of Alibaba \\
         Hangzhou, China}

\AddToShipoutPictureBG*{%
  \AtTextUpperLeft{%
    \put(\LenToUnit{\textwidth},\LenToUnit{8pt}){%
      \makebox(0,0)[br]{%
        \includegraphics[trim=7.5bp 9.5bp 10bp 9.5bp,clip,width=1.4in]{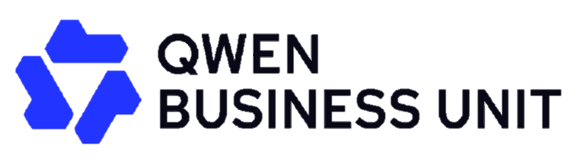}}}}}

\begin{document}

\maketitle

\begin{abstract}
Generative speech enhancement (SE) is prone to linguistic hallucination when semantic constraint is unreliable under severe noise and reverberation. Moreover, approaches that generate discrete codec tokens are bounded by the quantization error of the codec decoder, regardless of token-prediction accuracy. We propose LIFT-SE, a two-stage generative framework that decouples linguistic inference from acoustic synthesis within QRes-Codec, which exposes a quantized latent and its residual-completed continuous form. The first stage predicts clean codec tokens autoregressively, conditioned on frame-aligned features from a self-supervised front-end distilled toward clean speech. The second stage applies conditional flow matching to transport Gaussian noise to the continuous latent conditioned on the predicted tokens, and the frozen decoder reconstructs the enhanced waveform. Discrete generation provides naturalness, while continuous refinement restores signal fidelity. Experiments on the DNS1 and URGENT benchmarks show that LIFT-SE attains favorable linguistic consistency under reverberant conditions together with competitive perceptual quality, and systematic ablations verify the necessity of both stages. Code will be released in the future.
\end{abstract}

\begin{keywords}
Speech enhancement, autoregressive language model, conditional flow matching, neural audio codec, self-supervised feature distillation
\end{keywords}

\section{Introduction}

Speech enhancement (SE) aims to recover clean speech from noisy mixtures, improving perceptual quality and intelligibility for both human listeners and downstream applications such as automatic speech recognition (ASR) \cite{SE_for_ASR3} and speaker verification \cite{SE_for_SV2}. Deep-learning-based SE methods can be broadly grouped into discriminative approaches \cite{wang2023tfgridnet, yu2023efficient}, which estimate clean speech or its representations from noisy input, and generative approaches, which synthesize enhanced speech from learned speech priors. The latter include diffusion models \cite{lemercier2023storm, lu2022conditional}, flow matching models \cite{wang2025flowse, rong2026stupase}, and language-model (LM)-based approaches \cite{yao2025gense, kang2025llase, yan2025unise}. Generative SE can produce highly natural-sounding speech, but this capability entails a fundamental risk: perceptual plausibility does not guarantee content faithfulness, i.e., linguistic hallucination \cite{pirklbauer2023evaluation, rong2026pase}.

\looseness=-1 Achieving content faithfulness becomes particularly challenging when noise or reverberation obscures the acoustic evidence for the spoken words. Many LM-based SE systems generate discrete speech tokens, such as codec tokens, conditioned on representations of the degraded signal \cite{wang2024selm, yan2025unise}. When the conditioning is corrupted by severe noise or reverberation, the autoregressive prior dominates and the model produces fluent yet linguistically incorrect content \cite{wang2026l3se}. Continuous generative models similarly drift toward plausible but unfaithful speech in the absence of an explicit semantic anchor \cite{wang2025flowse, lemercier2023storm}. Existing work strengthens semantic guidance through noise-invariant representations: L3-SE distills noisy-speech representations toward clean acoustic and semantic teachers and reports strong linguistic consistency under reverberant conditions \cite{wang2026l3se}; PASE and StuPASE instead constrain generation in the continuous domain with a distilled self-supervised-learning (SSL) phonological prior \cite{rong2026pase, rong2026stupase}. However, the discrete modeling of L3-SE leaves reconstruction quality bounded by the quantization loss of the codec decoder, whereas the continuous-representation modeling of PASE and StuPASE falls short of autoregressive generation in naturalness. SenSE combines discrete token generation with flow-based waveform synthesis, but its token predictions and flow targets use different representations, requiring a mapping between the two stages \cite{li2025sense}. These observations motivate a framework that performs discrete linguistic inference and continuous acoustic refinement over a shared underlying speech representation.

\begin{figure*}[!t]
  \centering
  \includegraphics[width=\textwidth]{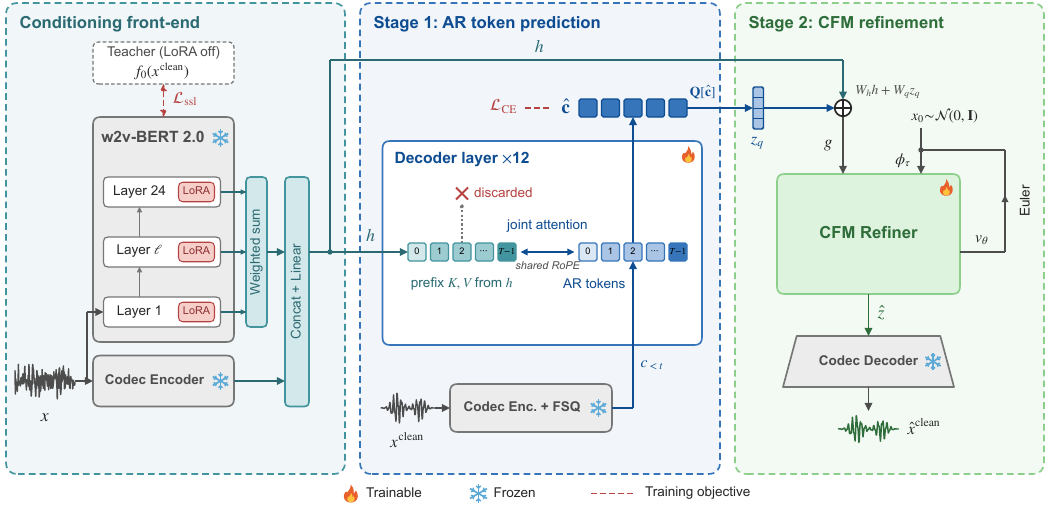}
  \caption{Overview of LIFT-SE. A LoRA-adapted self-supervised front-end and a frozen QRes-Codec encoder form the frame-aligned conditioning. Stage~1 predicts clean codec tokens, and Stage~2 refines their quantized latents into continuous latents before waveform reconstruction by the frozen QRes-Codec decoder.}
  \label{fig:overview}
\end{figure*}

We therefore propose LIFT-SE, a two-stage generative SE framework that decouples linguistic inference from acoustic synthesis. LIFT-SE is built on QRes-Codec, which exposes a quantized latent $z_q$ for discrete prediction and its residual-completed continuous form $z_q+z_{\mathrm{res}}$ for high-fidelity reconstruction. Concretely, the first stage is a conditional autoregressive (AR) model that predicts clean discrete codec tokens, conditioned on frame-aligned features from an SSL front-end and a frozen codec encoder that are injected as prepended key--value pairs at every decoder layer. The SSL front-end is adapted with low-rank adaptation (LoRA) and distilled toward the clean-speech representations of the same frozen backbone, which thus serves as its own teacher. The second stage is a conditional flow matching (CFM) refiner \cite{lipman2023flow} that transports Gaussian noise to the continuous latent of QRes-Codec, conditioned on the AR-predicted tokens; the frozen codec decoder then reconstructs the enhanced waveform.

Our contributions are as follows. First, we present QRes-Codec, in which discrete prediction and continuous refinement share a single representation space, eliminating the cross-representation conversion required when tokens and flow targets come from different representations. Second, we design a two-stage framework in which discrete autoregressive generation provides naturalness and an explicit causal decision over linguistic content, while continuous refinement recovers the information that quantization discards, lifting restoration quality beyond the fidelity of the discrete path. Third, on the DNS1 and URGENT benchmarks, LIFT-SE compares favorably with strong generative baselines under reverberant conditions while maintaining competitive perceptual quality. Removing either stage degrades the result, indicating that the two stages are complementary.


\section{Proposed method}
\label{sec:method}

\subsection{Overview}
\label{sec:method_overview}

\Cref{fig:overview} presents LIFT-SE, which enhances a noisy waveform $x$ in two stages that share a frozen QRes-Codec. QRes-Codec represents speech at a frame rate of $50$ Hz as a discrete token sequence $\mathbf{c}=(c_1,\ldots,c_T)$ and a continuous latent sequence. Let $\mathbf{Q}\in\mathbb{R}^{K\times 32}$ denote the dequantization table, where $K=4096$ is the codebook size. A table lookup maps tokens to quantized latents, $z_q=\mathbf{Q}[\mathbf{c}]$, allowing both stages to operate on representations defined by the same codec.

Stage~1 predicts clean codec tokens autoregressively from the noisy input and obtains $z_q$ by table lookup. Stage~2 uses conditional flow matching to estimate the target continuous latent $z=z_q+z_{\mathrm{res}}$, where $z_{\mathrm{res}}$ restores information omitted by quantization. The frozen codec decoder converts the refined latent $\hat{z}$ into the enhanced waveform $\hat{x}$. QRes-Codec and the self-supervised backbone (w2v-BERT-2.0) \cite{chung2021w2vbert} remain frozen, whereas the LoRA adapters, AR model, and CFM refiner are trained.

\subsection{QRes-Codec with semantic distillation}
\label{sec:method_codec}

\begin{figure}[!t]
  \centering
  \includegraphics[width=\linewidth]{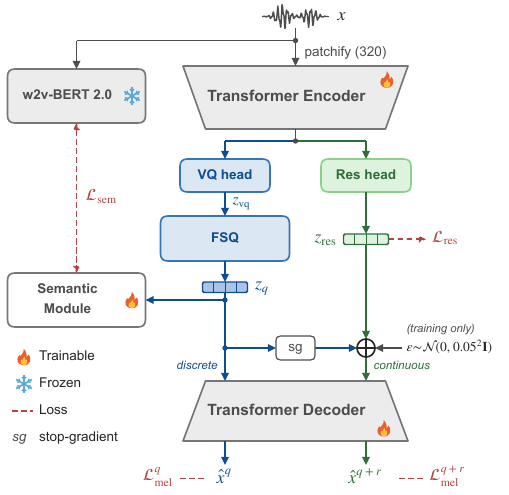}
  \caption{The frozen QRes-Codec provides a quantized latent $z_q$ for discrete prediction and a continuous latent $z_q+z_{\mathrm{res}}$ for high-fidelity reconstruction. Semantic distillation is applied to the quantized branch.}
  \label{fig:codec}
\end{figure}

Inspired by the homogeneous Transformer autoencoder design of MOSS-Audio-Tokenizer \cite{gong2026moss}, QRes-Codec (\Cref{fig:codec}) processes $16$-kHz waveforms using non-causal Transformer blocks. The waveform is divided into $320$-sample patches, yielding a frame rate of $50$ Hz; both the encoder and decoder contain $20$ Transformer layers with a hidden dimension of $768$. Two projection heads each map the encoder output to a $32$-dimensional latent. The first head produces $z_{\mathrm{vq}}$, which is quantized with finite scalar quantization (FSQ) \cite{mentzer2024finite} using levels $[4,4,4,4,4,4]$. The resulting single codebook contains $4^6=4096$ entries and has a bit rate of $600$ bps. The second head produces the residual $z_{\mathrm{res}}$. A single decoder reconstructs speech from $z_q$ in the discrete path and from $z_q+z_{\mathrm{res}}$ in the continuous path.

QRes-Codec is trained independently and then frozen for speech enhancement. Its objective combines reconstruction, residual, semantic, adversarial, and feature-matching terms:
\begin{align}
 \mathcal{L}_{\mathrm{codec}}={}&10\mathcal{L}_{\mathrm{mel}}^{q}
 +10\mathcal{L}_{\mathrm{mel}}^{q+r}
 +10\mathcal{L}_{\mathrm{res}}
 +5\mathcal{L}_{\mathrm{sem}} \nonumber\\
 &+\mathcal{L}_{\mathrm{adv}}+\mathcal{L}_{\mathrm{fm}},
 \label{eq:codec_loss}
\end{align}
where $\mathcal{L}_{\mathrm{mel}}^{q}$ and $\mathcal{L}_{\mathrm{mel}}^{q+r}$ denote multi-resolution log-mel losses for the discrete and continuous paths, respectively. During training, the quantized latent is detached in the continuous path, and $\mathcal{L}_{\mathrm{res}}=\lVert z_{\mathrm{res}}\rVert_2^2$ encourages the codes to retain the principal information. Multi-period and multi-resolution discriminators provide $\mathcal{L}_{\mathrm{adv}}$ and $\mathcal{L}_{\mathrm{fm}}$. Following semantic codec training \cite{li2025dualcodec}, a lightweight semantic module built on ConvNeXt~\cite{liu2022convnet} maps $z_q$ to the layer-17 representation of a frozen w2v-BERT-2.0 model \cite{chung2021w2vbert}; the mean-squared error defines $\mathcal{L}_{\mathrm{sem}}$. This loss encourages the quantized representation to retain linguistic information. Gaussian noise with a standard deviation of $0.05$ is added only to $z_q+z_{\mathrm{res}}$ during codec training, improving the robustness of the downstream refiner.

\subsection{Stage 1: conditional autoregressive token prediction}
\label{sec:method_ar}

\textbf{Conditioning front-end.}
We extract features from noisy speech $x$ using w2v-BERT-2.0, whose pretrained weights are frozen. Each of its $25$ hidden states is normalized by its own precomputed root-mean-square constant, and the normalized states are combined using learned softmax weights. The result is concatenated with the frame-aligned representation from the frozen codec encoder and projected to a sequence of fusion tokens $h\in\mathbb{R}^{T\times1024}$. Rank-$16$ LoRA adapters are added to the query, key, value, and output projections of each w2v-BERT-2.0 attention layer.

The LoRA adapters learn noise-invariant conditioning by aligning noisy-speech features with the clean-speech features of the same frozen backbone, following the noisy-to-clean representation distillation in PASE \cite{rong2026pase}. Let $f_{\mathrm{L}}$ denote the backbone with LoRA enabled and $f_0$ the backbone with LoRA disabled. The latter is evaluated without gradients and serves as the teacher:
\begin{equation}
 \mathcal{L}_{\mathrm{ssl}}=
 \left\lVert \mathrm{LN}\!\left(f_{\mathrm{L}}(x)\right)
 -\mathrm{LN}\!\left(f_0(x^{\mathrm{clean}})\right)\right\rVert_2^2.
 \label{eq:ssl_distill}
\end{equation}
Here, $\mathrm{LN}(\cdot)$ denotes layer normalization. Thus, the teacher exists but requires no additional training.

\textbf{Autoregressive model.}
The AR model is a decoder-only Transformer that follows the Qwen2 architecture \cite{yang2024qwen2}. It contains $12$ layers with a hidden dimension of $768$, $12$ attention heads, an intermediate dimension of $3072$, and a vocabulary of $4096$ codec tokens. No end-of-sequence token is needed because the output length equals the conditioning length. The model factorizes the clean token sequence as
\begin{equation}
 p_\theta(\mathbf{c}\mid h)=
 \prod_{t=1}^{T}p_\theta(c_t\mid c_{<t},h).
 \label{eq:ar_factorization}
\end{equation}

\textbf{Frame-aligned conditioning.}
At each decoder layer, an independent projection maps the fusion tokens $h$ to conditional key and value vectors, which are prepended to the keys and values derived from the AR-token states. After attention, outputs at the fusion-token positions are discarded, and only the outputs at the AR-token positions are propagated to the next layer. The fusion tokens and AR tokens share rotary indices $0,\ldots,T-1$; consequently, the two tokens at the same frame have zero relative rotary distance, independent of $T$. This framewise alignment allows inference to accommodate input audio of arbitrary duration.

\textbf{Objective and inference.}
Stage~1 is trained with teacher forcing under
\begin{equation}
 \mathcal{L}_{1}=\mathcal{L}_{\mathrm{CE}}
 +\mathcal{L}_{\mathrm{ssl}},
 \label{eq:stage1_loss}
\end{equation}
where $\mathcal{L}_{\mathrm{CE}}$ is the token cross-entropy. At inference, greedy generation produces $T$ tokens and the dequantization table maps them to $z_q$.

\subsection{Stage 2: conditional flow matching refinement}
\label{sec:method_cfm}

Stage~2 refines $z_q$ in the continuous codec representation rather than in a mel-spectrogram space. The $32$-dimensional codec latent is lower-dimensional than typical mel-spectrogram representations, and it can be decoded directly without a separately trained vocoder. Let $s_1=z=z_q^{\mathrm{clean}}+z_{\mathrm{res}}^{\mathrm{clean}}$ be the target continuous latent, where the superscript $\mathrm{clean}$ marks latents obtained from the corresponding clean speech, and let $s_0\sim\mathcal{N}(0,I)$ be the noise endpoint. For flow time $\tau\sim\mathcal{U}(0,1)$, we use the optimal-transport path \cite{lipman2023flow}
\begin{equation}
 \phi_\tau=(1-\tau)s_0+\tau s_1,
 \qquad u_\tau=s_1-s_0,
 \label{eq:ot_path}
\end{equation}
and train the velocity field with
\begin{equation}
 \mathcal{L}_{2}=
 \mathbb{E}_{s_0,s_1,\tau}
 \left[\left\lVert v_\theta(\phi_\tau,\tau\mid h,z_q)-u_\tau\right\rVert_2^2\right].
 \label{eq:cfm_loss}
\end{equation}
Here, $z_q$ is obtained from Stage~1, and the fusion tokens $h$ provide the noisy-speech evidence used by both stages.

The refiner consists of eight bidirectional Transformer blocks with a hidden dimension of $512$, eight attention heads, and an intermediate dimension of $2048$. Its condition is
\begin{equation}
 g=W_h h+W_q z_q.
 \label{eq:cfm_condition}
\end{equation}
The condition $g$ and $W_x\phi_\tau$ are concatenated along the time axis, and the two segments use the same frame indices for rotary position encoding. Here, $W_h$, $W_q$, and $W_x$ are learnable linear projections that map the fusion tokens $h$, quantized latent $z_q$, and interpolated latent $\phi_\tau$, respectively, to the refiner hidden dimension. The velocity prediction head is applied only to the outputs corresponding to the latent segment. A sinusoidal embedding of $\tau$ modulates every Transformer block through adaptive layer normalization \cite{peebles2023scalable}. During training, the entire condition is dropped with probability $0.2$ to enable classifier-free guidance \cite{ho2022classifier}.

At inference, Stage~1 first generates $\hat{\mathbf{c}}$ greedily, and the dequantization table yields $z_q=\mathbf{Q}[\hat{\mathbf{c}}]$. Starting from $s^{(0)}=s_0$, Euler integration updates
\begin{equation}
 s^{(i+1)}=s^{(i)}+\Delta\tau_i
 \left[v_{\mathrm{c}}+\gamma\left(v_{\mathrm{c}}-v_{\mathrm{u}}\right)\right],
 \label{eq:euler}
\end{equation}
where $v_{\mathrm{c}}$ and $v_{\mathrm{u}}$ are the velocities predicted with and without the condition, respectively, $\gamma$ is the guidance strength, and $\Delta\tau_i$ is the step size of the $i$-th integration step. The final latent $\hat{z}$ is passed to the frozen codec decoder to obtain $\hat{x}$.

\section{Experiments}
\label{sec:exp}

\subsection{Datasets}
\label{sec:datasets}

QRes-Codec is trained on VoxBox \cite{wang2025sparktts}, which contains approximately $100{,}000$ hours of speech. For SE training, clean speech is drawn from Multilingual LibriSpeech \cite{pratap2020mls} and Emilia \cite{he2024emilia}, about $3{,}000$ hours in total, and the noise collection, about $330$ hours in total, comprises the DNS Challenge noise set \cite{reddy2020interspeech}, DESED \cite{turpault2019desed}, DISCO \cite{furnon2021disco}, WHAM! \cite{wichern2019wham}, FSD50K \cite{fonseca2021fsd50k}, ESC-50 \cite{piczak2015esc}, DEMAND \cite{thiemann2013demand}, MUSAN \cite{snyder2015musan}, and TUT Urban Acoustic Scenes 2018 \cite{mesaros2018tut}. All training mixtures are generated online at $16$~kHz. We apply a simulated room impulse response (RIR) from SLR28 with probability $0.5$; the corresponding target is generated with the same RIR truncated $1$~ms after its direct-path peak. Noise is added with probability $0.8$ at an SNR sampled uniformly from $[-5,15]$~dB, after which the mixture is normalized to a level sampled from $[-30,-15]$~dB. We add three further distortions to the simulation to obtain a universal enhancement version, LIFT-SE-U: bandwidth limitation with probability $0.2$, which band-limits the signal to a sampling rate drawn from $\{3,4,6,8,11.025,12,16\}$~kHz; clipping with probability $0.2$, with the lower and upper thresholds set at sample quantiles drawn from $[0,0.1]$ and $[0.9,1]$; and packet loss with probability $0.15$, which drops $20$~ms packets at a loss rate drawn from $[0.05,0.25]$.

QRes-Codec is evaluated on the full LibriSpeech \textit{test-clean} split \cite{panayotov2015librispeech} ($2620$ utterances) as a clean-speech reconstruction task without simulated degradation. SE is evaluated on the synthetic DNS1 test set \cite{reddy2020interspeech}, which contains a \textit{no-reverb} and a \textit{with-reverb} subset of $150$ utterances each. For \textit{with-reverb}, the reference is the anechoic clean speech, so the evaluation covers dereverberation as well as denoising. LIFT-SE-U is additionally evaluated on the non-blind test set of the Interspeech 2025 URGENT challenge \cite{saijo2025interspeech}, which contains $1000$ multilingual utterances degraded by up to five of seven distortion types: additive noise, reverberation, clipping, bandwidth limitation, codec distortion, packet loss, and wind noise. The input sampling rates range from $8$ to $48$~kHz and are uniformly resampled to $16$~kHz for processing.

\subsection{Baselines}
\label{sec:baselines}

We compare QRes-Codec with EnCodec \cite{defossez2023high}, DAC \cite{kumar2023high}, X-Codec \cite{ye2025xcodec}, WavTokenizer \cite{ji2024wavtokenizer}, UniCodec \cite{jiang2025unicodec}, X-Codec2 \cite{ye2025llasa}, and BiCodec \cite{wang2025sparktts}. For SE, we include representative discriminative, continuous-domain, discrete-token, and hybrid systems. TF-GridNet \cite{wang2023tfgridnet} is evaluated using the URGENT 2025 baseline checkpoint \cite{saijo2025interspeech}. PASE \cite{rong2026pase}, UniPASE \cite{rong2026unipase}, and StuPASE \cite{rong2026stupase} use distilled self-supervised representations, with StuPASE performing CFM generation. UniSE \cite{yan2025unise} predicts discrete codec tokens autoregressively, and SenSE \cite{li2025sense} combines discrete semantic tokens with flow matching. We run all systems from their official checkpoints and apply the same evaluation pipeline to their outputs.

\begin{table}[!t]
  \centering
  \caption{QRes-Codec reconstruction on LibriSpeech \textit{test-clean}. FPS denotes frames per second, $N_q$ the number of codebooks, and BPS the bit rate in bits per second. Bold marks the best result excluding ground truth.}
  \label{tab:codec}
  \footnotesize
  \setlength{\tabcolsep}{2.6pt}
  \begin{adjustbox}{max width=\linewidth}
  \begin{tabular}{lcccccccc}
    \toprule
    Model & FPS & $N_q$ & BPS & PESQ$\uparrow$ & STOI$\uparrow$ & UTMOS$\uparrow$ & SIM$\uparrow$ & WER$\downarrow$ \\
    \midrule
    Ground truth & -- & -- & -- & 4.64 & 1.00 & 4.09 & 1.00 & 2.43 \\
    \midrule
    EnCodec      & 75 & 8 & 6000 & 2.77 & 0.94 & 3.09 & 0.89 & \textbf{2.64} \\
    DAC          & 50 & 4 & 2000 & 1.42 & 0.84 & 1.83 & 0.60 & 4.32 \\
    X-Codec      & 50 & 2 & 1000 & 2.33 & 0.86 & 4.21 & 0.72 & 3.13 \\
    WavTokenizer & 75 & 1 & 900  & 2.39 & 0.91 & 4.00 & 0.68 & 5.43 \\
    UniCodec     & 75 & 1 & 900  & 2.56 & 0.92 & 4.00 & 0.76 & 4.23 \\
    X-Codec2     & 50 & 1 & 800  & 2.43 & 0.92 & 4.13 & 0.82 & 3.53 \\
    BiCodec      & 50 & 1 & 650  & 2.51 & 0.92 & 4.18 & 0.80 & 3.23 \\
    QRes-Codec ($z_q$) & 50 & 1 & \textbf{600} & 2.22 & 0.91 & 4.19 & 0.69 & 5.64 \\
    \midrule
    QRes-Codec ($z_q+z_{\mathrm{res}}$) & 50 & -- & -- & \textbf{3.75} & \textbf{0.97} & \textbf{4.22} & \textbf{0.92} & 2.72 \\
    \bottomrule
  \end{tabular}
  \end{adjustbox}
\end{table}

\begin{table*}[!t]
  \centering
  \caption{Speech enhancement results on the DNS1 test set. Bold marks the best result.}
  \label{tab:main}
  \small
  \setlength{\tabcolsep}{3.5pt}
  \begin{adjustbox}{max width=\textwidth}
  \begin{tabular}{lcccccccccccc}
    \toprule
    & \multicolumn{6}{c}{DNS1 \textit{no-reverb}} & \multicolumn{6}{c}{DNS1 \textit{with-reverb}} \\
    \cmidrule(lr){2-7}\cmidrule(lr){8-13}
    Model & DNSMOS$\uparrow$ & UTMOS$\uparrow$ & SBS$\uparrow$ & LPS$\uparrow$ & SIM$\uparrow$ & dWER$\downarrow$
          & DNSMOS$\uparrow$ & UTMOS$\uparrow$ & SBS$\uparrow$ & LPS$\uparrow$ & SIM$\uparrow$ & dWER$\downarrow$ \\
    \midrule
    Noisy      & 2.48 & 2.36 & 0.88 & 0.90 & 0.96 & 3.81 & 1.39 & 1.30 & 0.74 & 0.63 & 0.79 & 10.99 \\
    \midrule
    TF-GridNet & 3.30 & 3.88 & 0.94 & \textbf{0.97} & \textbf{0.96} & 3.61 & 2.60 & 1.41 & 0.83 & 0.88 & 0.80 & 9.35 \\
    PASE       & 3.38 & 4.02 & 0.94 & \textbf{0.97} & 0.92 & 2.99 & 2.70 & 1.64 & 0.86 & 0.90 & 0.75 & 10.27 \\
    StuPASE    & 3.39 & 3.92 & 0.94 & \textbf{0.97} & 0.88 & \textbf{2.48} & 3.37 & 3.91 & 0.89 & 0.92 & 0.70 & 8.88 \\
    UniPASE    & 3.38 & 4.04 & \textbf{0.95} & \textbf{0.97} & 0.95 & 2.71 & 3.32 & 3.61 & \textbf{0.90} & \textbf{0.93} & 0.76 & 9.00 \\
    SenSE      & 3.39 & 3.85 & 0.94 & \textbf{0.97} & 0.92 & 5.78 & 3.38 & 3.56 & 0.88 & 0.92 & 0.76 & 11.72 \\
    UniSE      & \textbf{3.42} & \textbf{4.09} & 0.91 & 0.93 & 0.78 & 8.64 & \textbf{3.41} & 3.78 & 0.80 & 0.74 & 0.75 & 35.56 \\
    \midrule
    LIFT-SE    & 3.35 & 4.05 & \textbf{0.95} & \textbf{0.97} & 0.91 & 3.35 & 3.40 & \textbf{4.15} & \textbf{0.90} & \textbf{0.93} & \textbf{0.81} & \textbf{8.85} \\
    LIFT-SE-U & 3.37 & 4.07 & \textbf{0.95} & \textbf{0.97} & 0.92 & 4.00 & 3.40 & 4.12 & \textbf{0.90} & \textbf{0.93} & 0.80 & 8.94 \\
    \bottomrule
  \end{tabular}
  \end{adjustbox}
\end{table*}

\subsection{Evaluation metrics}
\label{sec:metrics}

We assess the outputs from four perspectives.
\begin{itemize}
\item \textbf{Perceptual quality}: DNSMOS P.835 \cite{reddy2022dnsmos} and UTMOS \cite{saeki2022utmos}, which predict perceived quality without a reference. We report the overall (OVRL) score of DNSMOS; because DNSMOS is sensitive to signal level, each enhanced output is scaled to the root-mean-square level of its noisy input before scoring.
\item \textbf{Signal fidelity}: wideband PESQ \cite{rec2001pesq} and STOI \cite{taal2011algorithm}, which compare the output with the reference waveform.
\item \textbf{Linguistic consistency}: SpeechBERTScore (SBS) \cite{saeki2024speechbertscore}, Levenshtein phoneme similarity (LPS) \cite{pirklbauer2023evaluation}, and word error rate, where SBS and LPS follow the URGENT 2025 challenge implementation \cite{saijo2025interspeech}. For QRes-Codec, WER is computed against the ground-truth text with a HuBERT-Large model.\footnote{\url{https://huggingface.co/facebook/hubert-large-ls960-ft}} DNS1 provides no transcripts, so for SE we report the differential WER (dWER) between the Whisper large-v3 \cite{radford2023robust} transcripts of the enhanced speech and of the clean reference. For URGENT 2025, we report the WER between the transcripts of the output speech and the reference text.
\item \textbf{Speaker preservation}: speaker similarity (SIM), the cosine similarity between speaker embeddings of the output and the reference, extracted by a speaker verification model based on WavLM-Large features \cite{chen2022wavlm}.\footnote{\url{https://github.com/microsoft/UniSpeech/tree/main/downstreams/speaker_verification}}
\end{itemize}
For QRes-Codec, we report PESQ, STOI, UTMOS, SIM, and WER; for SE, we report DNSMOS, UTMOS, SBS, LPS, SIM, dWER, and WER. WER and dWER are reported as percentages. Higher is better for all metrics except WER and dWER.

\subsection{Implementation details}
\label{sec:impl}

QRes-Codec contains approximately $307$M parameters and is trained with AdamW using a learning rate of $1.8\times10^{-4}$ for both the generator and the discriminators and gradient clipping at $5.0$; it remains frozen in all SE experiments. The $580$M-parameter w2v-BERT~2.0 backbone is also frozen, while rank-$16$ LoRA adapters with $\alpha=32$ and dropout $0.05$ are applied to the query, key, value, and output projections of every attention layer, adding approximately $3.1$M trainable parameters. The $140$M-parameter AR model is trained for $80$k updates using AdamW with a peak learning rate of $3\times10^{-4}$. The $49.5$M-parameter CFM refiner is trained for $50$k updates with the same optimizer settings while Stage~1 is frozen, with token conditions sampled from teacher-forced Stage~1 predictions. At inference, Stage~1 decodes greedily, and the refiner uses $4$ Euler steps ($6$ for LIFT-SE-U) on a sway-sampled time grid with coefficient $-1$ \cite{chen2025f5tts} and guidance strength $\gamma=0.5$. LIFT-SE contains approximately $193$M trainable parameters.
\subsection{QRes-Codec evaluation}
\label{sec:codec_eval}

\Cref{tab:codec} compares the discrete and continuous reconstruction paths of QRes-Codec. At $600$~bps, the quantized path obtains a UTMOS of $4.19$, but its PESQ, SIM, and WER indicate a substantial loss of fine acoustic detail, speaker characteristics, and linguistic content. Since UTMOS is a learned non-intrusive predictor, values above that of the ground-truth recordings should not be interpreted as higher reconstruction fidelity \cite{liu2025unitok}.

The continuous path retains similar predicted naturalness while improving PESQ, STOI, SIM, and WER to $3.75$, $0.97$, $0.92$, and $2.72$, respectively, indicating that $z_{\mathrm{res}}$ contains acoustic information critical to improving fidelity. The contrast between the two paths motivates the proposed decomposition: Stage~1 predicts the discrete representation, and Stage~2 recovers the information absent from that representation.

\subsection{Speech enhancement results}
\label{sec:main_results}

\Cref{tab:main} compares the systems on DNS1. On \textit{no-reverb}, LIFT-SE performs competitively in predicted naturalness and phonetic preservation. Under reverberant conditions, LIFT-SE attains the lowest dWER together with the highest UTMOS and speaker similarity, exhibiting a favorable balance across naturalness, speaker similarity, and linguistic fidelity. In contrast, the high DNSMOS and UTMOS of UniSE do not translate into comparable dWER, underscoring the need to assess generative SE with both perceptual and content-based metrics. LIFT-SE-U performs comparably to LIFT-SE on both subsets, indicating that training on a wider range of distortions does not degrade performance in conventional scenarios. Spectrograms and transcripts of selected \textit{with-reverb} utterances are provided in Appendix~\ref{sec:appendix}.

\subsection{\texorpdfstring{Evaluation on URGENT 2025}{Evaluation on URGENT 2025}}
\label{sec:universal}

\begin{table}[!t]
  \centering
  \caption{Results on the URGENT 2025 non-blind test set. Bold marks the best result among the enhancement systems.}
  \label{tab:urgent}
  \footnotesize
  \setlength{\tabcolsep}{3.5pt}
  \begin{tabular}{lcccccc}
    \toprule
    Model & DNSMOS$\uparrow$ & UTMOS$\uparrow$ & SBS$\uparrow$ & LPS$\uparrow$ & SIM$\uparrow$ & WER$\downarrow$ \\
    \midrule
    Noisy      & 1.83 & 1.56 & 0.75 & 0.62 & 0.81 & 27.99 \\
    \midrule
    TF-GridNet & 2.96 & 2.13 & 0.86 & 0.79 & \textbf{0.86} & 27.67 \\
    UniPASE    & 3.21 & 3.00 & \textbf{0.89} & \textbf{0.84} & 0.83 & 27.78 \\
    UniSE      & \textbf{3.26} & 3.08 & 0.78 & 0.60 & 0.58 & 54.90 \\
    \midrule
    LIFT-SE-U  & 3.24 & \textbf{3.22} & 0.87 & \textbf{0.84} & 0.81 & \textbf{23.11} \\
    \bottomrule
  \end{tabular}
\end{table}

We evaluate LIFT-SE-U on the URGENT 2025 non-blind test set to assess its performance on universal speech enhancement. As shown in \cref{tab:urgent}, all enhancement systems raise DNSMOS and UTMOS well above the noisy input. UniSE obtains the highest DNSMOS, but its WER is very high, whereas TF-GridNet and UniPASE keep the WER close to that of the noisy input. Although LIFT-SE-U sees no codec distortion or wind noise during training, it still obtains the best UTMOS, WER, and LPS. However, its SIM is lower than that of TF-GridNet, a purely discriminative model, possibly because generative modeling tends to restore natural-sounding speech under more complex distortions at the cost of some speaker characteristics.

\subsection{Ablation study}
\label{sec:ablation}

\begin{table*}[!t]
  \centering
  \caption{Ablations on DNS1. Stage~1 variants are decoded directly from $z_q$ without the refiner and are compared with the Stage~1-only system; refiner variants are compared with LIFT-SE. Bold marks the best result.}
  \label{tab:ablation}
  \small
  \setlength{\tabcolsep}{3.5pt}
  \begin{adjustbox}{max width=\textwidth}
  \begin{tabular}{lcccccccccccc}
    \toprule
    & \multicolumn{6}{c}{DNS1 \textit{no-reverb}} & \multicolumn{6}{c}{DNS1 \textit{with-reverb}} \\
    \cmidrule(lr){2-7}\cmidrule(lr){8-13}
    Variant & DNSMOS$\uparrow$ & UTMOS$\uparrow$ & SBS$\uparrow$ & LPS$\uparrow$ & SIM$\uparrow$ & dWER$\downarrow$
            & DNSMOS$\uparrow$ & UTMOS$\uparrow$ & SBS$\uparrow$ & LPS$\uparrow$ & SIM$\uparrow$ & dWER$\downarrow$ \\
    \midrule
    LIFT-SE (CFM refiner)       & 3.35 & 4.05 & \textbf{0.95} & \textbf{0.97} & \textbf{0.91} & \textbf{3.35} & 3.40 & 4.15 & \textbf{0.90} & \textbf{0.93} & \textbf{0.81} & 8.85 \\
    \midrule
    Stage~1 only                    & 3.42 & 4.19 & 0.92 & 0.94 & 0.73 & 6.72  & 3.40 & 4.20 & 0.88 & 0.91 & 0.67 & 11.28 \\
    \quad w/o $\mathcal{L}_{\mathrm{ssl}}$ & \textbf{3.43} & 4.19 & 0.92 & 0.93 & 0.73 & 7.77  & 3.40 & 4.16 & 0.88 & 0.91 & 0.66 & 12.21 \\
    \quad w/o LoRA                  & 3.42 & 4.19 & 0.91 & 0.93 & 0.72 & 8.34  & 3.38 & 4.09 & 0.85 & 0.84 & 0.61 & 21.42 \\
    \quad w/o SSL features          & \textbf{3.43} & 4.16 & 0.88 & 0.88 & 0.69 & 17.06 & 3.39 & 3.94 & 0.80 & 0.71 & 0.52 & 43.64 \\
    \quad w/o codec-encoder features & 3.40 & 4.22 & 0.91 & 0.94 & 0.72 & 6.76  & 3.38 & 4.17 & 0.88 & 0.91 & 0.66 & 11.29 \\
    \midrule
    CFM $\rightarrow$ regression refiner & 3.42 & \textbf{4.25} & \textbf{0.95} & \textbf{0.97} & 0.90 & 3.61 & \textbf{3.43} & \textbf{4.24} & \textbf{0.90} & \textbf{0.93} & 0.77 & 9.11 \\
    trained on generated tokens & 3.33 & 4.05 & \textbf{0.95} & \textbf{0.97} & \textbf{0.91} & 3.52 & 3.39 & 4.08 & \textbf{0.90} & \textbf{0.93} & 0.80 & \textbf{8.63} \\
    w/o $z_q$ (no Stage~1)     & 3.34 & 4.04 & \textbf{0.95} & \textbf{0.97} & \textbf{0.91} & 4.08 & 3.36 & 4.08 & \textbf{0.90} & \textbf{0.93} & 0.80 & 9.22 \\
    \bottomrule
  \end{tabular}
  \end{adjustbox}
\end{table*}

\Cref{tab:ablation} isolates the design choices of each stage. Comparing Stage~1-only decoding with LIFT-SE shows that continuous refinement improves speaker similarity and linguistic fidelity while retaining high predicted naturalness, supporting the intended division of labor between the two stages.

\textbf{Stage~1 conditioning.} Removing the SSL features causes the largest degradation across linguistic, phonetic, and speaker-related metrics, particularly under reverberation. Removing LoRA also substantially weakens linguistic consistency, whereas removing the distillation loss and the codec-encoder features produces a smaller but consistent decline. These trends show that the SSL representation provides the main linguistic constraint and that noisy-to-clean adaptation makes this constraint more robust to reverberation.

\textbf{Refiner.} The regression refiner obtains higher DNSMOS and UTMOS, whereas CFM yields higher speaker similarity and a lower dWER on both subsets. Removing the $z_q$ condition raises the dWER, reflecting the contribution of Stage~1 to linguistic fidelity. Training the refiner on generated tokens, i.e., on tokens that Stage~1 generates autoregressively as at inference instead of teacher-forced predictions, yields the lowest \textit{with-reverb} dWER among the refiner variants, accompanied by slight reductions in predicted perceptual quality. This result suggests that exposure to Stage~1 generation errors may help the refiner correct some of these errors under reverberation at the cost of slightly reduced naturalness.

\begin{figure}[!t]
  \centering
  \includegraphics[width=\linewidth]{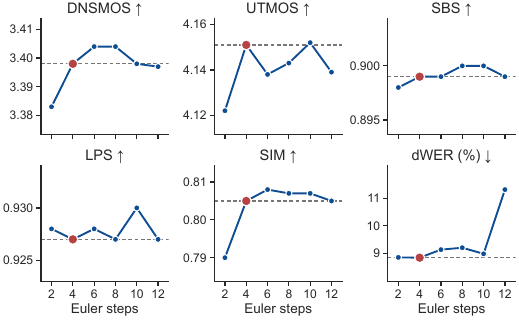}
  \caption{Effect of the number of Euler steps on DNS1 \textit{with-reverb} for LIFT-SE. The red marker and the dashed line denote the default four steps.}
  \label{fig:flowsteps}
\end{figure}

\textbf{Number of flow steps.} 
As shown in \Cref{fig:flowsteps}, all metrics change little with the number of Euler steps. 
Therefore, we utilize four steps to provide a good balance between quality and inference cost.

\section{Conclusion}
\label{sec:conclusion}

We presented LIFT-SE, a two-stage generative SE system that separates discrete linguistic inference from continuous acoustic refinement within the representation space of QRes-Codec. Evaluations of QRes-Codec and LIFT-SE consistently show that autoregressive token prediction primarily establishes speech naturalness, while refinement in the continuous latent improves signal fidelity, speaker preservation, and linguistic consistency. On DNS1, LIFT-SE is particularly competitive under reverberation, and the ablations support the complementary roles of the two stages. Extending the training simulation to more distortion types largely maintains this behavior on DNS1 and yields a favorable word error rate on the multi-distortion URGENT 2025 non-blind test set. Some limitations remain: sequential autoregressive decoding increases inference latency, and speaker similarity in complex scenarios still has room for improvement. Future work will consider improved speaker preservation and more efficient token generation.

\bibliographystyle{IEEEtran}
\bibliography{refs}

@article{wang2026l3se,
  author    = {Wang, Zheng and Rong, Xiaobin and Su, Hang and Tan, Tianyi and Wu, Junnan and Fan, Lichun and Luo, Zhenbo and Luan, Jian and Lu, Jing},
  title     = {Reducing Linguistic Hallucination in {LM}-Based Speech Enhancement via Noise-Invariant Acoustic-Semantic Distillation},
  journal   = {arXiv preprint arXiv:2605.08608},
  year      = {2026},
  url       = {https://arxiv.org/abs/2605.08608}
}

@inproceedings{rong2026pase,
  author    = {Rong, Xiaobin and Hu, Qinwen and Yesilbursa, Mansur and Wojcicki, Kamil and Lu, Jing},
  title     = {{PASE}: Leveraging the Phonological Prior of {WavLM} for Low-Hallucination Generative Speech Enhancement},
  booktitle = {Proceedings of the AAAI Conference on Artificial Intelligence},
  volume    = {40},
  number    = {39},
  pages     = {32826--32834},
  year      = {2026},
  doi       = {10.1609/aaai.v40i39.40562},
  url       = {https://arxiv.org/abs/2511.13300}
}

@article{rong2026stupase,
  author    = {Rong, Xiaobin and Gao, Jun and Wang, Zheng and Yesilbursa, Mansur and Wojcicki, Kamil and Lu, Jing},
  title     = {{StuPASE}: Towards Low-Hallucination Studio-Quality Generative Speech Enhancement},
  journal   = {arXiv preprint arXiv:2603.09234},
  year      = {2026},
  url       = {https://arxiv.org/abs/2603.09234}
}

@article{li2025sense,
  author    = {Li, Xingchen and Xie, Hanke and Wang, Ziqian and Zhang, Zihan and Xiao, Longshuai and Wang, Shuai and Xie, Lei},
  title     = {{SenSE}: Semantic-Aware High-Fidelity Universal Speech Enhancement},
  journal   = {arXiv preprint arXiv:2509.24708},
  year      = {2025},
  url       = {https://arxiv.org/abs/2509.24708},
  note      = {Accepted to ICME 2026 (arXiv v2 comment).}
}

@inproceedings{wang2025flowse,
  author    = {Wang, Ziqian and Liu, Zikai and Zhu, Xinfa and Zhu, Yike and Liu, Mingshuai and Chen, Jun and Xiao, Longshuai and Weng, Chao and Xie, Lei},
  title     = {{FlowSE}: Efficient and High-Quality Speech Enhancement via Flow Matching},
  booktitle = {Proceedings of Interspeech},
  pages     = {4858--4862},
  year      = {2025},
  doi       = {10.21437/Interspeech.2025-1745},
  note      = {Wang et al., Interspeech 2025. Distinct from Lee et al., ICASSP 2025, which uses the same short name FlowSE.}
}

@inproceedings{yao2025gense,
  author    = {Yao, Jixun and Liu, Hexin and Chen, Chen and Hu, Yuchen and Chng, Eng Siong and Xie, Lei},
  title     = {{GenSE}: Generative Speech Enhancement via Language Models using Hierarchical Modeling},
  booktitle = {The Thirteenth International Conference on Learning Representations (ICLR)},
  year      = {2025},
  url       = {https://openreview.net/forum?id=1p6xFLBU4J}
}

@article{yan2025unise,
  author    = {Yan, Haoyin and Liu, Chengwei and Xue, Shaofei and Liang, Xiaotao and Liu, Yinghao and Kong, Yuxiang and Xue, Zheng},
  title     = {{UniSE}: A Unified Framework for Decoder-Only Autoregressive {LM}-Based Speech Enhancement},
  journal   = {arXiv preprint arXiv:2510.20441},
  year      = {2025},
  url       = {https://arxiv.org/abs/2510.20441},
  note      = {Accepted to Interspeech 2026 (arXiv v2 comment).}
}

@inproceedings{kang2025llase,
  author    = {Kang, Boyi and Zhu, Xinfa and Zhang, Zihan and Ye, Zhen and Liu, Mingshuai and Wang, Ziqian and Zhu, Yike and Ma, Guobin and Chen, Jun and Xiao, Longshuai and Weng, Chao and Xue, Wei and Xie, Lei},
  title     = {{LLaSE-G1}: Incentivizing Generalization Capability for {LLaMA}-based Speech Enhancement},
  booktitle = {Proceedings of the 63rd Annual Meeting of the Association for Computational Linguistics (Volume 1: Long Papers)},
  pages     = {13292--13305},
  year      = {2025},
  address   = {Vienna, Austria},
  doi       = {10.18653/v1/2025.acl-long.651}
}

@inproceedings{wang2024selm,
  author    = {Wang, Ziqian and Zhu, Xinfa and Zhang, Zihan and Lv, Yuanjun and Jiang, Ning and Zhao, Guoqing and Xie, Lei},
  title     = {{SELM}: Speech Enhancement Using Discrete Tokens and Language Models},
  booktitle = {IEEE International Conference on Acoustics, Speech and Signal Processing (ICASSP)},
  pages     = {11561--11565},
  year      = {2024},
  doi       = {10.1109/ICASSP48485.2024.10447464}
}

@article{lemercier2023storm,
  author    = {Lemercier, Jean-Marie and Richter, Julius and Welker, Simon and Gerkmann, Timo},
  title     = {{StoRM}: A Diffusion-Based Stochastic Regeneration Model for Speech Enhancement and Dereverberation},
  journal   = {IEEE/ACM Transactions on Audio, Speech, and Language Processing},
  volume    = {31},
  pages     = {2724--2737},
  year      = {2023},
  doi       = {10.1109/TASLP.2023.3294692}
}

@inproceedings{lu2022conditional,
  author    = {Lu, Yen-Ju and Wang, Zhong-Qiu and Watanabe, Shinji and Richard, Alexander and Yu, Cheng and Tsao, Yu},
  title     = {Conditional Diffusion Probabilistic Model for Speech Enhancement},
  booktitle = {IEEE International Conference on Acoustics, Speech and Signal Processing (ICASSP)},
  pages     = {7402--7406},
  year      = {2022},
  doi       = {10.1109/ICASSP43922.2022.9746901}
}

@article{defossez2023high,
  author    = {D{\'e}fossez, Alexandre and Copet, Jade and Synnaeve, Gabriel and Adi, Yossi},
  title     = {High Fidelity Neural Audio Compression},
  journal   = {Transactions on Machine Learning Research (TMLR)},
  year      = {2023},
  url       = {https://openreview.net/forum?id=ivCd8z8zR2}
}

@inproceedings{kumar2023high,
  author    = {Kumar, Rithesh and Seetharaman, Prem and Luebs, Alejandro and Kumar, Ishaan and Kumar, Kundan},
  title     = {High-Fidelity Audio Compression with Improved {RVQGAN}},
  booktitle = {Advances in Neural Information Processing Systems (NeurIPS)},
  volume    = {36},
  pages     = {27980--27993},
  year      = {2023},
  doi       = {10.52202/075280-1214}
}

@inproceedings{ji2024wavtokenizer,
  author    = {Ji, Shengpeng and Jiang, Ziyue and Wang, Wen and Chen, Yifu and Fang, Minghui and Zuo, Jialong and Yang, Qian and Cheng, Xize and Wang, Zehan and Li, Ruiqi and Zhang, Ziang and Yang, Xiaoda and Huang, Rongjie and Jiang, Yidi and Chen, Qian and Zheng, Siqi and Zhao, Zhou},
  title     = {{WavTokenizer}: An Efficient Acoustic Discrete Codec Tokenizer for Audio Language Modeling},
  booktitle = {The Thirteenth International Conference on Learning Representations (ICLR)},
  year      = {2025},
  url       = {https://openreview.net/forum?id=yBlVlS2Fd9}
}

@article{ye2025llasa,
  author    = {Ye, Zhen and Zhu, Xinfa and Chan, Chi-Min and Wang, Xinsheng and Tan, Xu and Lei, Jiahe and Peng, Yi and Liu, Haohe and Jin, Yizhu and Dai, Zheqi and Lin, Hongzhan and Chen, Jianyi and Du, Xingjian and Xue, Liumeng and Chen, Yunlin and Li, Zhifei and Xie, Lei and Kong, Qiuqiang and Guo, Yike and Xue, Wei},
  title     = {Llasa: Scaling Train-Time and Inference-Time Compute for {LLaMA}-based Speech Synthesis},
  journal   = {arXiv preprint arXiv:2502.04128},
  year      = {2025},
  url       = {https://arxiv.org/abs/2502.04128}
}

@article{wang2025sparktts,
  author    = {Wang, Xinsheng and Jiang, Mingqi and Ma, Ziyang and Zhang, Ziyu and Liu, Songxiang and Li, Linqin and Liang, Zheng and Zheng, Qixi and Wang, Rui and Feng, Xiaoqin and Bian, Weizhen and Ye, Zhen and Cheng, Sitong and Yuan, Ruibin and Zhao, Zhixian and Zhu, Xinfa and Pan, Jiahao and Xue, Liumeng and Zhu, Pengcheng and Chen, Yunlin and Li, Zhifei and Chen, Xie and Xie, Lei and Guo, Yike and Xue, Wei},
  title     = {{Spark-TTS}: An Efficient {LLM}-Based Text-to-Speech Model with Single-Stream Decoupled Speech Tokens},
  journal   = {arXiv preprint arXiv:2503.01710},
  year      = {2025},
  url       = {https://arxiv.org/abs/2503.01710}
}

@inproceedings{lipman2023flow,
  author    = {Lipman, Yaron and Chen, Ricky T. Q. and Ben-Hamu, Heli and Nickel, Maximilian and Le, Matt},
  title     = {Flow Matching for Generative Modeling},
  booktitle = {The Eleventh International Conference on Learning Representations (ICLR)},
  year      = {2023},
  url       = {https://openreview.net/forum?id=PqvMRDCJT9t}
}

@inproceedings{peebles2023scalable,
  author    = {Peebles, William and Xie, Saining},
  title     = {Scalable Diffusion Models with Transformers},
  booktitle = {IEEE/CVF International Conference on Computer Vision (ICCV)},
  pages     = {4172--4182},
  year      = {2023},
  doi       = {10.1109/ICCV51070.2023.00387}
}

@inproceedings{chen2025f5tts,
  author    = {Chen, Yushen and Niu, Zhikang and Ma, Ziyang and Deng, Keqi and Wang, Chunhui and Zhao, Jian and Yu, Kai and Chen, Xie},
  title     = {{F5-TTS}: A Fairytaler that Fakes Fluent and Faithful Speech with Flow Matching},
  booktitle = {Proceedings of the 63rd Annual Meeting of the Association for Computational Linguistics (Volume 1: Long Papers)},
  pages     = {6255--6271},
  year      = {2025},
  address   = {Vienna, Austria},
  doi       = {10.18653/v1/2025.acl-long.313}
}

@article{chen2022wavlm,
  author    = {Chen, Sanyuan and Wang, Chengyi and Chen, Zhengyang and Wu, Yu and Liu, Shujie and Chen, Zhuo and Li, Jinyu and Kanda, Naoyuki and Yoshioka, Takuya and Xiao, Xiong and others},
  title     = {{WavLM}: Large-Scale Self-Supervised Pre-Training for Full Stack Speech Processing},
  journal   = {IEEE Journal of Selected Topics in Signal Processing},
  volume    = {16},
  number    = {6},
  pages     = {1505--1518},
  year      = {2022},
  doi       = {10.1109/JSTSP.2022.3188113}
}

@inproceedings{saijo2025interspeech,
  author    = {Saijo, Kohei and Zhang, Wangyou and Cornell, Samuele and Scheibler, Robin and Li, Chenda and Ni, Zhaoheng and Kumar, Anurag and Sach, Marvin and Fu, Yihui and Wang, Wei and Fingscheidt, Tim and Watanabe, Shinji},
  title     = {Interspeech 2025 {URGENT} Speech Enhancement Challenge},
  booktitle = {Proceedings of Interspeech},
  pages     = {858--862},
  year      = {2025},
  doi       = {10.21437/Interspeech.2025-1363}
}

@inproceedings{reddy2020interspeech,
  author    = {Reddy, Chandan K. A. and Gopal, Vishak and Cutler, Ross and Beyrami, Ebrahim and Cheng, Roger and Dubey, Harishchandra and Matusevych, Sergiy and Aichner, Robert and Aazami, Ashkan and Braun, Sebastian and Rana, Puneet and Srinivasan, Sriram and Gehrke, Johannes},
  title     = {The {INTERSPEECH} 2020 Deep Noise Suppression Challenge: Datasets, Subjective Testing Framework, and Challenge Results},
  booktitle = {Proceedings of Interspeech},
  pages     = {2492--2496},
  year      = {2020},
  doi       = {10.21437/Interspeech.2020-3038}
}

@inproceedings{panayotov2015librispeech,
  author    = {Panayotov, Vassil and Chen, Guoguo and Povey, Daniel and Khudanpur, Sanjeev},
  title     = {{LibriSpeech}: An {ASR} Corpus Based on Public Domain Audio Books},
  booktitle = {IEEE International Conference on Acoustics, Speech and Signal Processing (ICASSP)},
  pages     = {5206--5210},
  year      = {2015},
  doi       = {10.1109/ICASSP.2015.7178964}
}

@inproceedings{wichern2019wham,
  author    = {Wichern, Gordon and Antognini, Joe and Flynn, Michael and Zhu, Licheng Richard and McQuinn, Emmett and Crow, Dwight and Manilow, Ethan and Le Roux, Jonathan},
  title     = {{WHAM!}: Extending Speech Separation to Noisy Environments},
  booktitle = {Proceedings of Interspeech},
  pages     = {1368--1372},
  year      = {2019},
  doi       = {10.21437/Interspeech.2019-2821}
}

@article{fonseca2021fsd50k,
  author    = {Fonseca, Eduardo and Favory, Xavier and Pons, Jordi and Font, Frederic and Serra, Xavier},
  title     = {{FSD50K}: An Open Dataset of Human-Labeled Sound Events},
  journal   = {IEEE/ACM Transactions on Audio, Speech, and Language Processing},
  volume    = {30},
  pages     = {829--852},
  year      = {2022},
  doi       = {10.1109/TASLP.2021.3133208}
}

@inproceedings{turpault2019desed,
  author    = {Turpault, Nicolas and Serizel, Romain and Parag Shah, Ankit and Salamon, Justin},
  title     = {Sound Event Detection in Domestic Environments with Weakly Labeled Data and Soundscape Synthesis},
  booktitle = {Proceedings of the Detection and Classification of Acoustic Scenes and Events Workshop (DCASE)},
  pages     = {253--257},
  year      = {2019},
  doi       = {10.33682/006b-jx26}
}

@article{furnon2021disco,
  author    = {Furnon, Nicolas and Serizel, Romain and Essid, Slim and Illina, Irina},
  title     = {{DNN}-Based Mask Estimation for Distributed Speech Enhancement in Spatially Unconstrained Microphone Arrays},
  journal   = {IEEE/ACM Transactions on Audio, Speech, and Language Processing},
  volume    = {29},
  pages     = {2310--2323},
  year      = {2021},
  doi       = {10.1109/TASLP.2021.3092838}
}

@inproceedings{piczak2015esc,
  author    = {Piczak, Karol J.},
  title     = {{ESC}: Dataset for Environmental Sound Classification},
  booktitle = {Proceedings of the 23rd ACM International Conference on Multimedia},
  pages     = {1015--1018},
  year      = {2015},
  doi       = {10.1145/2733373.2806390}
}

@inproceedings{thiemann2013demand,
  author    = {Thiemann, Joachim and Ito, Nobutaka and Vincent, Emmanuel},
  title     = {The Diverse Environments Multi-Channel Acoustic Noise Database ({DEMAND}): A Database of Multichannel Environmental Noise Recordings},
  booktitle = {Proceedings of Meetings on Acoustics},
  volume    = {19},
  number    = {1},
  pages     = {035081},
  year      = {2013},
  doi       = {10.1121/1.4799597}
}

@article{snyder2015musan,
  author    = {Snyder, David and Chen, Guoguo and Povey, Daniel},
  title     = {{MUSAN}: A Music, Speech, and Noise Corpus},
  journal   = {arXiv preprint arXiv:1510.08484},
  year      = {2015},
  url       = {https://arxiv.org/abs/1510.08484}
}

@inproceedings{mesaros2018tut,
  author    = {Mesaros, Annamaria and Heittola, Toni and Virtanen, Tuomas},
  title     = {A Multi-Device Dataset for Urban Acoustic Scene Classification},
  booktitle = {Proceedings of the Detection and Classification of Acoustic Scenes and Events Workshop (DCASE)},
  pages     = {9--13},
  year      = {2018},
  url       = {https://arxiv.org/abs/1807.09840}
}

@inproceedings{he2024emilia,
  author    = {He, Haorui and Shang, Zengqiang and Wang, Chaoren and Li, Xuyuan and Gu, Yicheng and Hua, Hua and Liu, Liwei and Yang, Chen and Li, Jiaqi and Shi, Peiyang and others},
  title     = {Emilia: An Extensive, Multilingual, and Diverse Speech Dataset for Large-Scale Speech Generation},
  booktitle = {IEEE Spoken Language Technology Workshop (SLT)},
  pages     = {885--890},
  year      = {2024},
  doi       = {10.1109/SLT61566.2024.10832365}
}

@inproceedings{reddy2022dnsmos,
  author    = {Reddy, Chandan K. A. and Gopal, Vishak and Cutler, Ross},
  title     = {{DNSMOS P.835}: A Non-Intrusive Perceptual Objective Speech Quality Metric to Evaluate Noise Suppressors},
  booktitle = {IEEE International Conference on Acoustics, Speech and Signal Processing (ICASSP)},
  pages     = {886--890},
  year      = {2022},
  doi       = {10.1109/ICASSP43922.2022.9746108}
}

@inproceedings{saeki2022utmos,
  author    = {Saeki, Takaaki and Xin, Detai and Nakata, Wataru and Koriyama, Tomoki and Takamichi, Shinnosuke and Saruwatari, Hiroshi},
  title     = {{UTMOS}: {UTokyo-SaruLab} System for {VoiceMOS Challenge 2022}},
  booktitle = {Proceedings of Interspeech},
  pages     = {4521--4525},
  year      = {2022},
  doi       = {10.21437/Interspeech.2022-439}
}

@inproceedings{saeki2024speechbertscore,
  author    = {Saeki, Takaaki and Maiti, Soumi and Takamichi, Shinnosuke and Watanabe, Shinji and Saruwatari, Hiroshi},
  title     = {{SpeechBERTScore}: Reference-Aware Automatic Evaluation of Speech Generation Leveraging {NLP} Evaluation Metrics},
  booktitle = {Proceedings of Interspeech},
  pages     = {4943--4947},
  year      = {2024},
  doi       = {10.21437/Interspeech.2024-1508}
}

@inproceedings{pirklbauer2023evaluation,
  author    = {Pirklbauer, Jan and Sach, Marvin and Fluyt, Kristoff and Tirry, Wouter and Wardah, Wafaa and Moeller, Sebastian and Fingscheidt, Tim},
  title     = {Evaluation Metrics for Generative Speech Enhancement Methods: Issues and Perspectives},
  booktitle = {Speech Communication; 15th ITG Conference},
  pages     = {265--269},
  year      = {2023}
}

@inproceedings{radford2023robust,
  author    = {Radford, Alec and Kim, Jong Wook and Xu, Tao and Brockman, Greg and Mcleavey, Christine and Sutskever, Ilya},
  title     = {Robust Speech Recognition via Large-Scale Weak Supervision},
  booktitle = {International Conference on Machine Learning (ICML)},
  volume    = {202},
  pages     = {28492--28518},
  year      = {2023}
}

@article{wang2023tfgridnet,
  author    = {Wang, Zhong-Qiu and Cornell, Samuele and Choi, Shukjae and Lee, Younglo and Kim, Byeong-Yeol and Watanabe, Shinji},
  title     = {{TF-GridNet}: Integrating Full- and Sub-Band Modeling for Speech Separation},
  journal   = {IEEE/ACM Transactions on Audio, Speech, and Language Processing},
  volume    = {31},
  pages     = {3221--3236},
  year      = {2023},
  doi       = {10.1109/TASLP.2023.3304482}
}

@inproceedings{yu2023efficient,
  author    = {Yu, Jianwei and Luo, Yi},
  title     = {Efficient Monaural Speech Enhancement with Universal Sample Rate Band-Split {RNN}},
  booktitle = {IEEE International Conference on Acoustics, Speech and Signal Processing (ICASSP)},
  pages     = {1--5},
  year      = {2023},
  doi       = {10.1109/ICASSP49357.2023.10096020}
}

@misc{rec2001pesq,
  author       = {{ITU-T Recommendation P.862}},
  title        = {Perceptual Evaluation of Speech Quality ({PESQ}): An Objective Method for End-to-End Speech Quality Assessment of Narrow-Band Telephone Networks and Speech Codecs},
  year         = {2001},
  howpublished = {International Telecommunication Union}
}

@inproceedings{mentzer2024finite,
  author    = {Mentzer, Fabian and Minnen, David and Agustsson, Eirikur and Tschannen, Michael},
  title     = {Finite Scalar Quantization: {VQ-VAE} Made Simple},
  booktitle = {The Twelfth International Conference on Learning Representations ({ICLR})},
  year      = {2024},
  url       = {https://openreview.net/forum?id=8ishA3LxN8}
}

@inproceedings{SE_for_ASR3,
  title     = {End-to-End Integration of Speech Recognition, Speech Enhancement, and Self-Supervised Learning Representation},
  author    = {Xuankai Chang and Takashi Maekaku and Yuya Fujita and Shinji Watanabe},
  year      = {2022},
  booktitle = {Interspeech 2022},
  pages     = {3819--3823}
}

@inproceedings{SE_for_SV2,
  title     = {A Framework for Robust Speaker Verification in Highly Noisy Environments Leveraging Both Noisy and Enhanced Audio},
  author    = {Katav, Adam and Moshe, Yair and Cohen, Israel},
  booktitle = {2025 33rd European Signal Processing Conference (EUSIPCO)},
  pages     = {41--45},
  year      = {2025},
  organization= {IEEE}
}

@inproceedings{li2025dualcodec,
  author    = {Li, Jiaqi and Lin, Xiaolong and Li, Zhekai and Huang, Shixi and Wang, Yuancheng and Wang, Chaoren and Zhan, Zhenpeng and Wu, Zhizheng},
  title     = {{DualCodec}: A Low-Frame-Rate, Semantically-Enhanced Neural Audio Codec for Speech Generation},
  booktitle = {Interspeech 2025},
  year      = {2025},
  url       = {https://arxiv.org/abs/2505.13000}
}

@inproceedings{liu2022convnet,
  author    = {Liu, Zhuang and Mao, Hanzi and Wu, Chao-Yuan and Feichtenhofer, Christoph and Darrell, Trevor and Xie, Saining},
  title     = {A {ConvNet} for the 2020s},
  booktitle = {Proceedings of the IEEE/CVF Conference on Computer Vision and Pattern Recognition (CVPR)},
  pages     = {11976--11986},
  year      = {2022},
  doi       = {10.1109/CVPR52688.2022.01167}
}

@inproceedings{chung2021w2vbert,
  author    = {Chung, Yu-An and Zhang, Yu and Han, Wei and Chiu, Chung-Cheng and Qin, James and Pang, Ruoming and Wu, Yonghui},
  title     = {{w2v-BERT}: Combining Contrastive Learning and Masked Language Modeling for Self-Supervised Speech Pre-Training},
  booktitle = {IEEE Automatic Speech Recognition and Understanding Workshop (ASRU)},
  year      = {2021},
  doi       = {10.1109/ASRU51503.2021.9688253}
}

@article{yang2024qwen2,
  author  = {Yang, An and Yang, Baosong and Hui, Binyuan and Zheng, Bo and Yu, Bowen and Zhou, Chang and Li, Chengpeng and Li, Chengyuan and Liu, Dayiheng and Huang, Fei and others},
  title   = {{Qwen2} Technical Report},
  journal = {arXiv preprint arXiv:2407.10671},
  year    = {2024},
  url     = {https://arxiv.org/abs/2407.10671}
}

@article{gong2026moss,
  author  = {Gong, Yitian and Chen, Kuangwei and Fei, Zhaoye and Yang, Xiaogui and Chen, Ke and Wang, Yang and Huang, Kexin and Chen, Mingshu and Li, Ruixiao and Cheng, Qingyuan and Li, Shimin and Qiu, Xipeng},
  title   = {{MOSS-Audio-Tokenizer}: Scaling Audio Tokenizers for Future Audio Foundation Models},
  journal = {arXiv preprint arXiv:2602.10934},
  year    = {2026},
  url     = {https://arxiv.org/abs/2602.10934}
}

@inproceedings{ho2022classifier,
  author    = {Ho, Jonathan and Salimans, Tim},
  title     = {Classifier-Free Diffusion Guidance},
  booktitle = {NeurIPS 2021 Workshop on Deep Generative Models and Downstream Applications},
  year      = {2021},
  url       = {https://arxiv.org/abs/2207.12598}
}

@inproceedings{ye2025xcodec,
  author    = {Ye, Zhen and Sun, Peiwen and Lei, Jiahe and Lin, Hongzhan and Tan, Xu and Dai, Zheqi and Kong, Qiuqiang and Chen, Jianyi and Pan, Jiahao and Liu, Qifeng and Guo, Yike and Xue, Wei},
  title     = {Codec Does Matter: Exploring the Semantic Shortcoming of Codec for Audio Language Model},
  booktitle = {Proceedings of the AAAI Conference on Artificial Intelligence},
  volume    = {39},
  number    = {24},
  pages     = {25697--25705},
  year      = {2025},
  doi       = {10.1609/aaai.v39i24.34761},
  note      = {X-Codec.}
}

@inproceedings{jiang2025unicodec,
  author    = {Jiang, Yidi and Chen, Qian and Ji, Shengpeng and Xi, Yu and Wang, Wen and Zhang, Chong and Yue, Xianghu and Zhang, ShiLiang and Li, Haizhou},
  title     = {{UniCodec}: Unified Audio Codec with Single Domain-Adaptive Codebook},
  booktitle = {Proceedings of the 63rd Annual Meeting of the Association for Computational Linguistics (Volume 1: Long Papers)},
  pages     = {19112--19124},
  year      = {2025},
  doi       = {10.18653/v1/2025.acl-long.937}
}

@article{liu2025unitok,
  author    = {Liu, Chengwei and Yan, Haoyin and Xue, Shaofei and Liang, Xiaotao and Liu, Yinghao and Xue, Zheng and Song, Gang and Zhou, Boyang},
  title     = {{UniTok-Audio}: A Unified Audio Generation Framework via Generative Modeling on Discrete Codec Tokens},
  journal   = {arXiv preprint arXiv:2510.26372},
  year      = {2025},
  url       = {https://arxiv.org/abs/2510.26372},
  note      = {Source of the codec baseline numbers (Table 2).}
}

@article{rong2026unipase,
  author    = {Rong, Xiaobin and Wang, Zheng and Wang, Yushi and Gao, Jun and Lu, Jing},
  title     = {{UniPASE}: A Generative Model for Universal Speech Enhancement With High Fidelity and Low Hallucinations},
  journal   = {IEEE Transactions on Audio, Speech and Language Processing},
  volume    = {34},
  pages     = {3901--3915},
  year      = {2026},
  doi       = {10.1109/TASLPRO.2026.3717231}
}

@article{taal2011algorithm,
  author    = {Taal, Cees H. and Hendriks, Richard C. and Heusdens, Richard and Jensen, Jesper},
  title     = {An Algorithm for Intelligibility Prediction of Time--Frequency Weighted Noisy Speech},
  journal   = {IEEE Transactions on Audio, Speech, and Language Processing},
  volume    = {19},
  number    = {7},
  pages     = {2125--2136},
  year      = {2011},
  doi       = {10.1109/TASL.2011.2114881},
  note      = {STOI.}
}

@inproceedings{pratap2020mls,
  author    = {Pratap, Vineel and Xu, Qiantong and Sriram, Anuroop and Synnaeve, Gabriel and Collobert, Ronan},
  title     = {{MLS}: A Large-Scale Multilingual Dataset for Speech Research},
  booktitle = {Proc. Interspeech},
  pages     = {2757--2761},
  year      = {2020},
  doi       = {10.21437/Interspeech.2020-2826}
}

\onecolumn
\appendix
\section{Spectrogram comparison on DNS1 with-reverb samples}
\label{sec:appendix}
\renewcommand{\thefigure}{A\arabic{figure}}
\renewcommand{\theHfigure}{A\arabic{figure}}
\setcounter{figure}{0}
\definecolor{cardbg}{HTML}{F8FAFC}
\definecolor{cardborder}{HTML}{CBD5E1}
\definecolor{oursborder}{HTML}{22C55E}
\definecolor{errred}{HTML}{D00000}
\newcommand{\spectrocard}[5]{%
  {\setlength{\fboxsep}{3pt}\setlength{\fboxrule}{0.9pt}%
   \fcolorbox{#5}{cardbg}{\parbox[t]{\dimexpr0.318\textwidth-2\fboxsep-2\fboxrule\relax}{%
     \centering{\footnotesize\textbf{#1}}\\[2pt]
     \includegraphics[width=\linewidth]{#2}\\[2pt]
     \raggedright\scriptsize #3\\[1pt] #4}}}}

\begin{figure*}[!ht]
  \centering
  \footnotesize
  \noindent
    \spectrocard{Noisy}{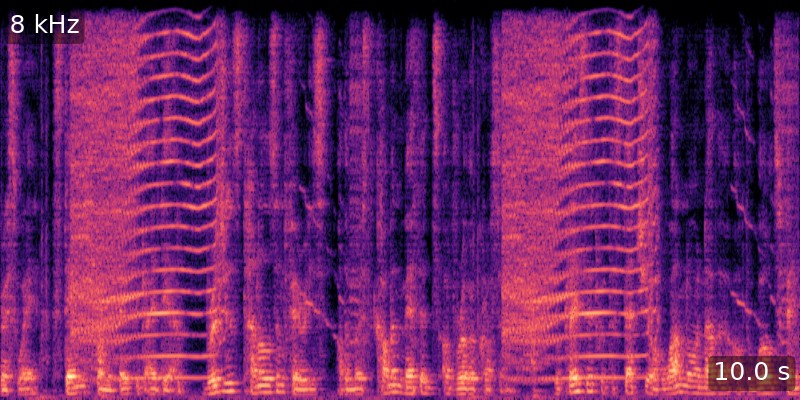}{WER 9.7\% (S2, D1, I0)}{\textcolor{errred}{stays} from the basic idea bridges tunnels and ferries are the most common methods of river \textcolor{errred}{crossing} replace it with the statue of one or another of the worlds famous}{cardborder}
    \hfill
    \spectrocard{TF-GridNet}{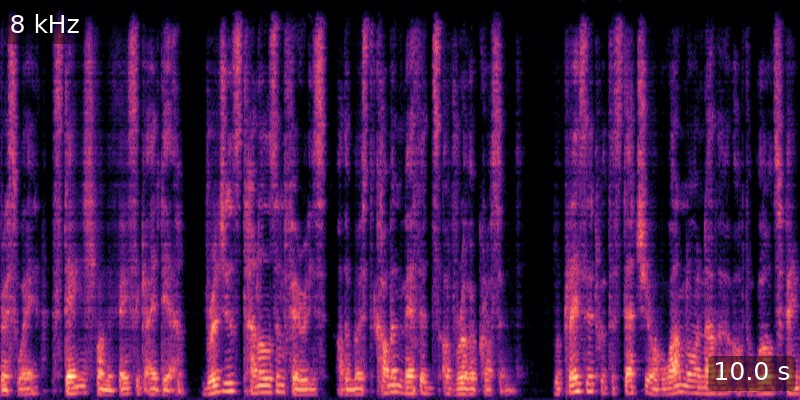}{WER 9.7\% (S2, D1, I0)}{\textcolor{errred}{stange} from the basic idea bridges tunnels and ferries are the most common methods of river \textcolor{errred}{crossing} replace it with the statue of one or another of the worlds famous}{cardborder}
    \hfill
    \spectrocard{PASE}{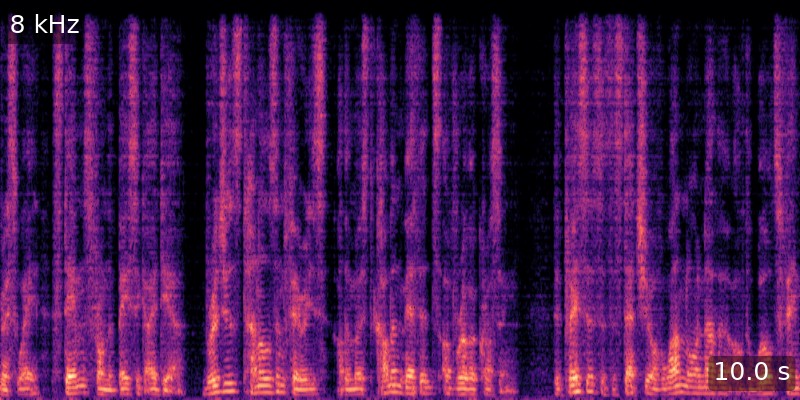}{WER 16.1\% (S5, D0, I0)}{\textcolor{errred}{railway} stems from the basic idea bridges tunnels and ferries are the most common methods of river \textcolor{errred}{crossing} \textcolor{errred}{the} \textcolor{errred}{places} \textcolor{errred}{is} the statue of one or another of the worlds famous}{cardborder}\par\smallskip
  \noindent
    \spectrocard{StuPASE}{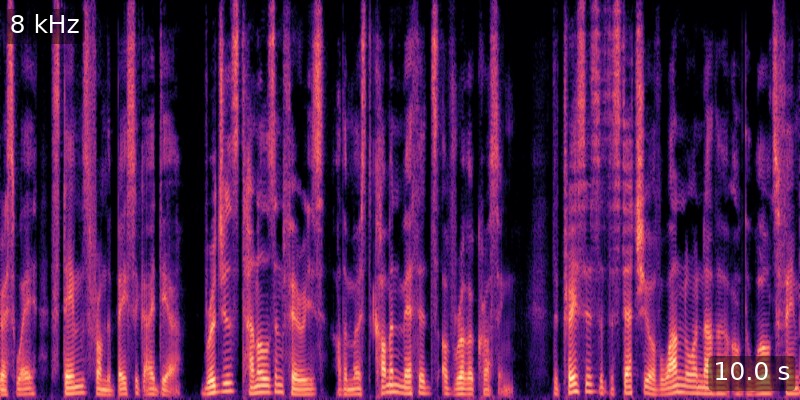}{WER 12.9\% (S4, D0, I0)}{way stems from the basic idea bridges tunnels and ferries are the most common methods of river \textcolor{errred}{crossing} \textcolor{errred}{the} \textcolor{errred}{traces} \textcolor{errred}{of} the statue of one or another of the worlds famous}{cardborder}
    \hfill
    \spectrocard{UniPASE}{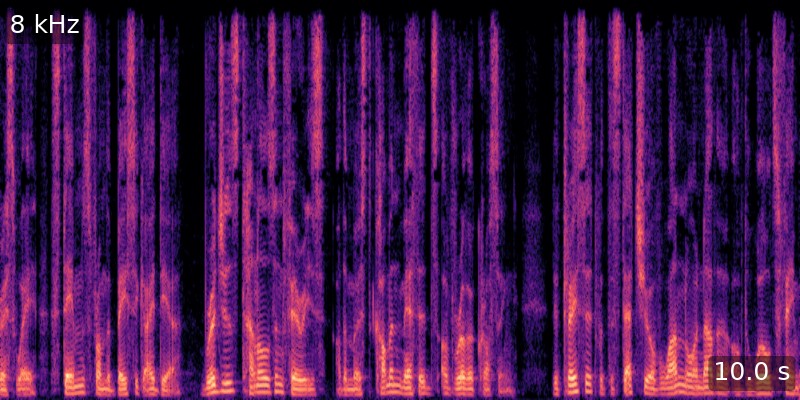}{WER 9.7\% (S2, D0, I1)}{way stems from the basic idea bridges tunnels and ferries are the most common methods of river \textcolor{errred}{crossing} \textcolor{errred}{they} \textcolor{errred}{trace} it with the statue of one or another of the worlds famous}{cardborder}
    \hfill
    \spectrocard{SenSE}{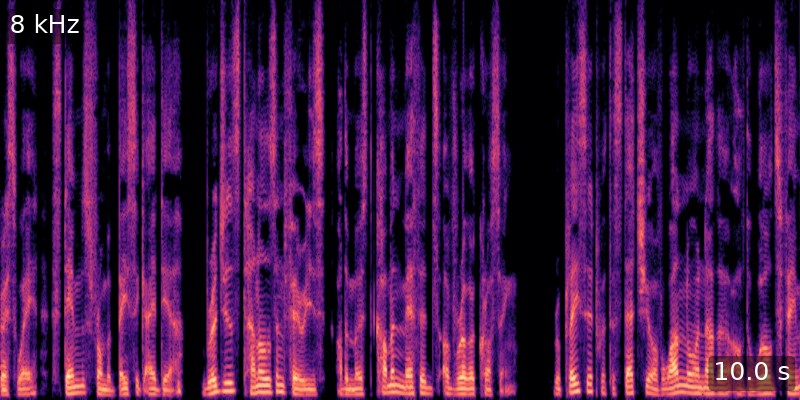}{WER 9.7\% (S3, D0, I0)}{way stems from \textcolor{errred}{a} basic idea bridges tunnels and ferries are the most common methods of river \textcolor{errred}{crossing} replace it with \textcolor{errred}{a} statue of one or another of the worlds famous}{cardborder}\par\smallskip
  \noindent
    \spectrocard{UniSE}{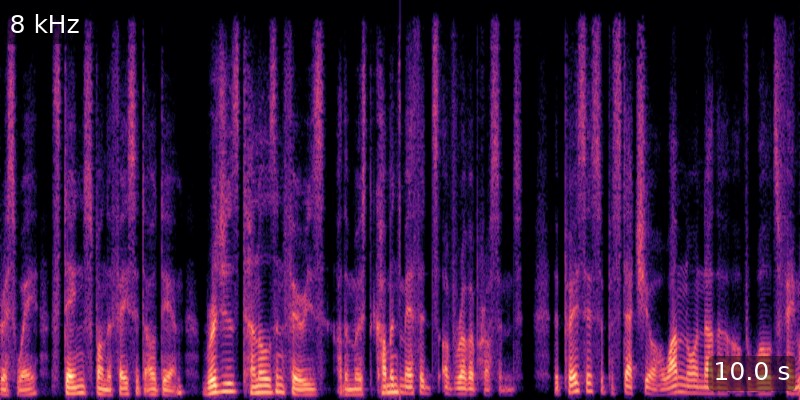}{WER 38.7\% (S5, D4, I3)}{\textcolor{errred}{railway} \textcolor{errred}{stains} from the \textcolor{errred}{face} \textcolor{errred}{at} \textcolor{errred}{our} \textcolor{errred}{dam} bridges tunnels and ferries are the most common methods of river crossings the \textcolor{errred}{process} of \textcolor{errred}{pistachio} one or another of the worlds}{cardborder}
    \hfill
    \spectrocard{LIFT-SE (Ours)}{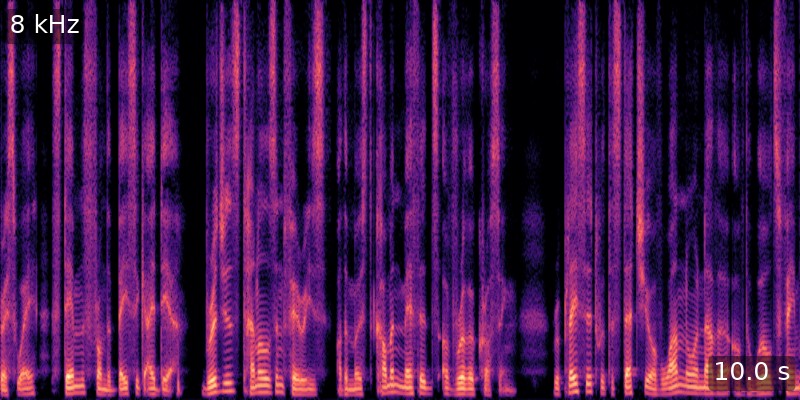}{WER 3.2\% (S1, D0, I0)}{way stems from the basic idea bridges tunnels and ferries are the most common methods of river \textcolor{errred}{crossing} replace it with the statue of one or another of the worlds famous}{oursborder}
    \hfill
    \spectrocard{Clean}{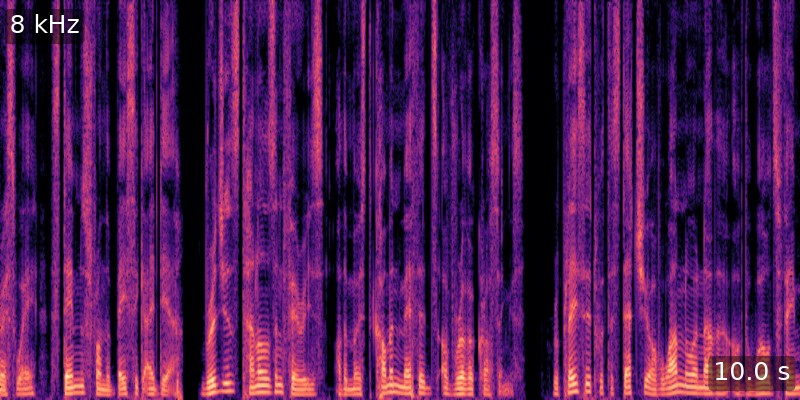}{Reference transcript}{way stems from the basic idea bridges tunnels and ferries are the most common methods of river crossings replace it with the statue of one or another of the worlds famous}{cardborder}
  \caption{Spectrograms and Whisper large-v3 transcripts of DNS1 \textit{with-reverb} utterance \texttt{fileid\_76}.}
  \label{fig:cards_1}
\end{figure*}

\begin{figure*}[!ht]
  \centering
  \footnotesize
  \noindent
    \spectrocard{Noisy}{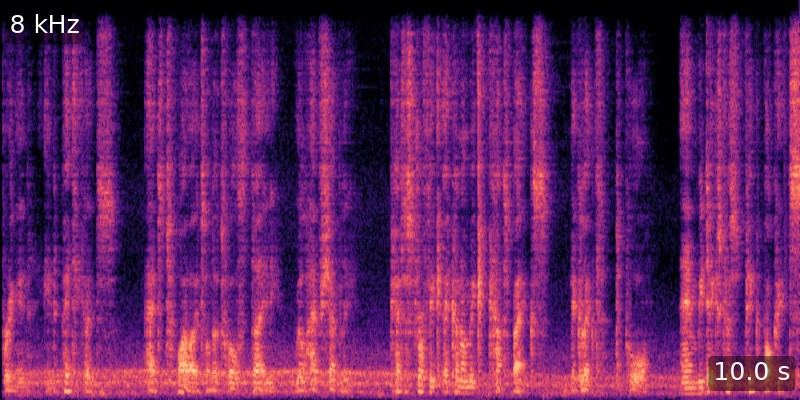}{WER 9.5\% (S2, D0, I0)}{bring in in petticoats at twilight on the twelfth day well have \textcolor{errred}{shadley} catastrophic economic cutbacks neglect \textcolor{errred}{to} poor ambidextrous pickpockets}{cardborder}
    \hfill
    \spectrocard{TF-GridNet}{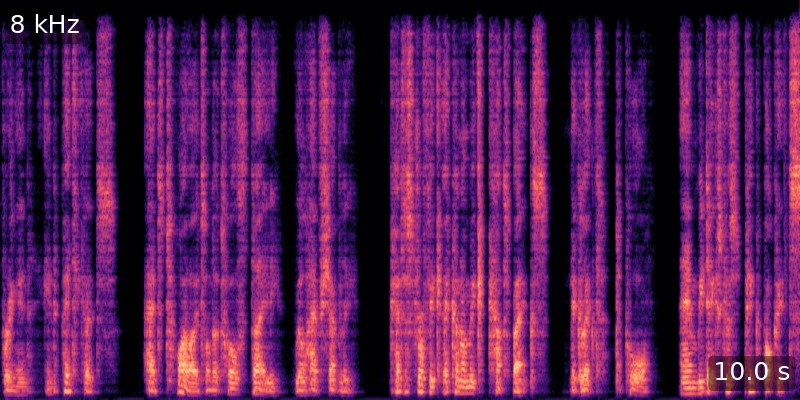}{WER 9.5\% (S2, D0, I0)}{bring in in petticoats \textcolor{errred}{add} twilight on the twelfth day well have \textcolor{errred}{shadley} catastrophic economic cutbacks neglect the poor ambidextrous pickpockets}{cardborder}
    \hfill
    \spectrocard{PASE}{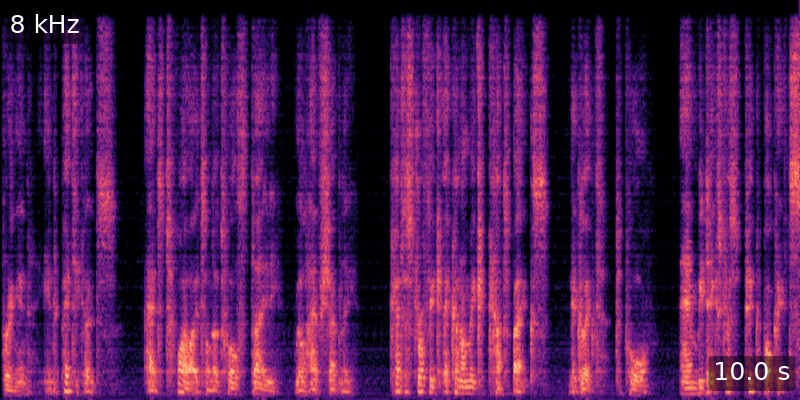}{WER 9.5\% (S2, D0, I0)}{bring in in petticoats at twilight on the twelfth day well have \textcolor{errred}{shadley} catastrophic economic cutbacks neglect \textcolor{errred}{to} poor ambidextrous pickpockets}{cardborder}\par\smallskip
  \noindent
    \spectrocard{StuPASE}{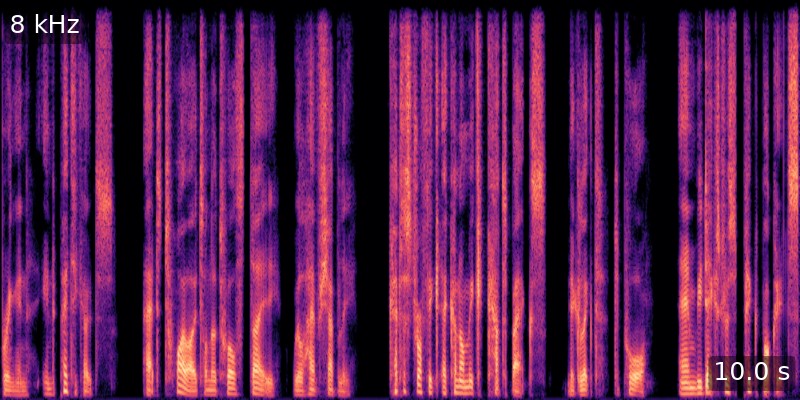}{WER 9.5\% (S0, D2, I0)}{in petticoats at twilight on the twelfth day well have shabbily catastrophic economic cutbacks neglect the poor ambidextrous pickpockets}{cardborder}
    \hfill
    \spectrocard{UniPASE}{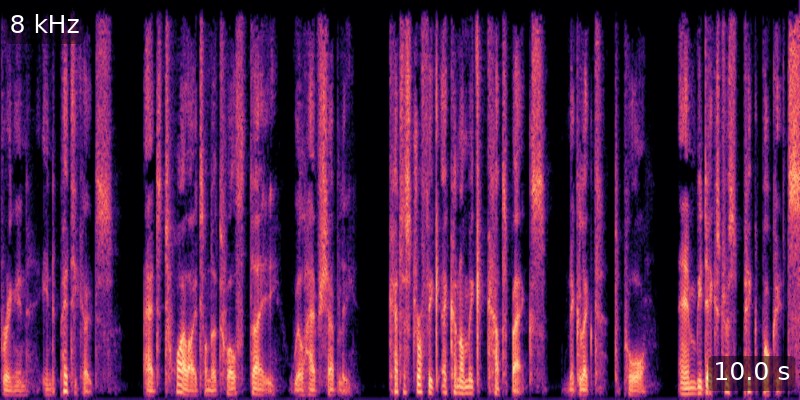}{WER 9.5\% (S2, D0, I0)}{bring in in petticoats \textcolor{errred}{add} twilight on the twelfth day well have shabbily catastrophic economic cutbacks neglect \textcolor{errred}{to} poor ambidextrous pickpockets}{cardborder}
    \hfill
    \spectrocard{SenSE}{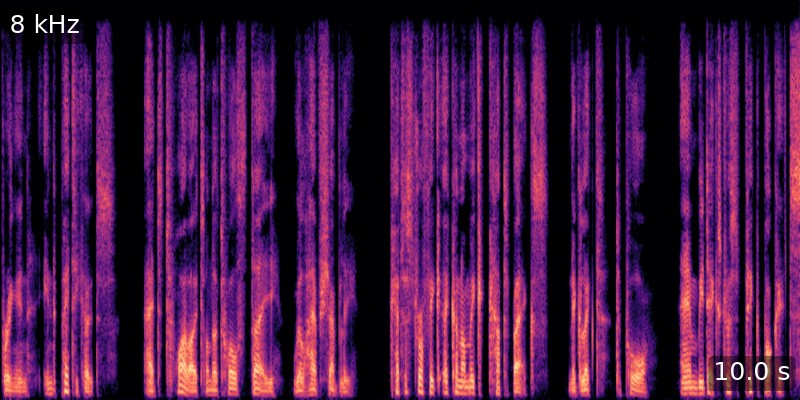}{WER 14.3\% (S2, D1, I0)}{\textcolor{errred}{springing} in petticoats at twilight on the twelfth day well have shabbily catastrophic economic cutbacks neglect \textcolor{errred}{to} poor ambidextrous pickpockets}{cardborder}\par\smallskip
  \noindent
    \spectrocard{UniSE}{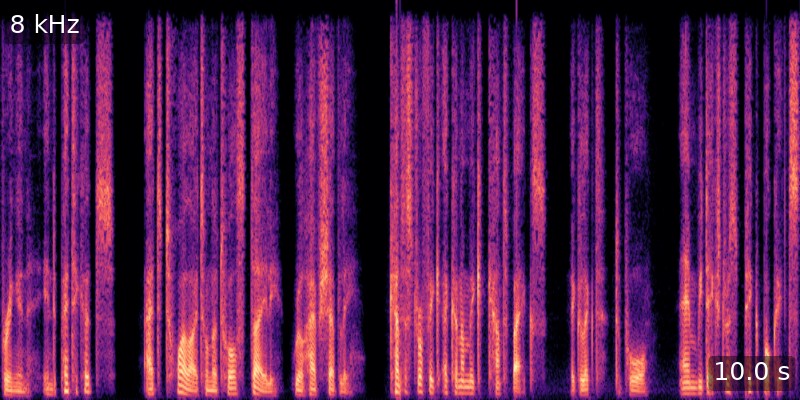}{WER 23.8\% (S4, D0, I1)}{bring in in \textcolor{errred}{the} petticoats at twilight on the twelfth \textcolor{errred}{daily} well have \textcolor{errred}{sheveley} catastrophic economic cutbacks \textcolor{errred}{eclect} \textcolor{errred}{to} poor ambidextrous pickpockets}{cardborder}
    \hfill
    \spectrocard{LIFT-SE (Ours)}{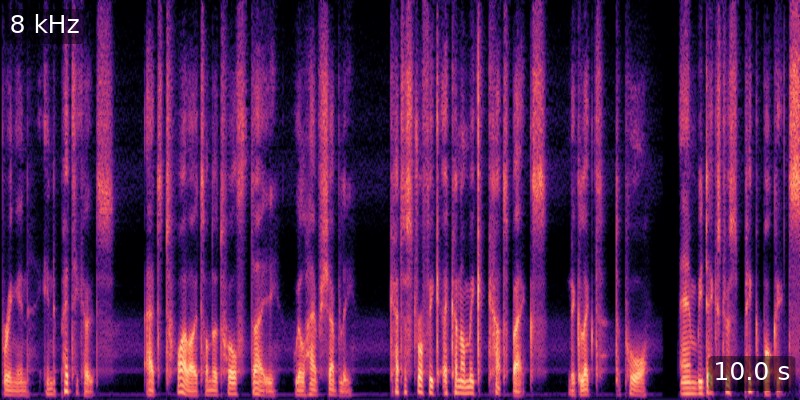}{WER 4.8\% (S1, D0, I0)}{bring in in petticoats at twilight on the twelfth day well have shabbily catastrophic economic cutbacks neglect \textcolor{errred}{to} poor ambidextrous pickpockets}{oursborder}
    \hfill
    \spectrocard{Clean}{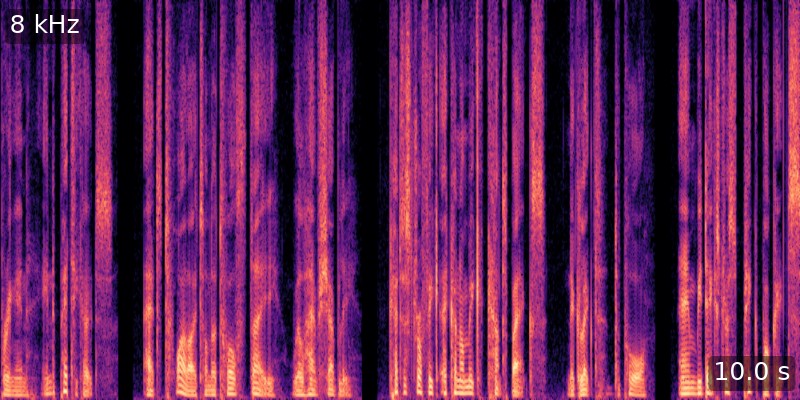}{Reference transcript}{bring in in petticoats at twilight on the twelfth day well have shabbily catastrophic economic cutbacks neglect the poor ambidextrous pickpockets}{cardborder}
  \caption{Spectrograms and Whisper large-v3 transcripts of DNS1 \textit{with-reverb} utterance \texttt{fileid\_96}.}
  \label{fig:cards_2}
\end{figure*}

\begin{figure*}[!ht]
  \centering
  \footnotesize
  \noindent
    \spectrocard{Noisy}{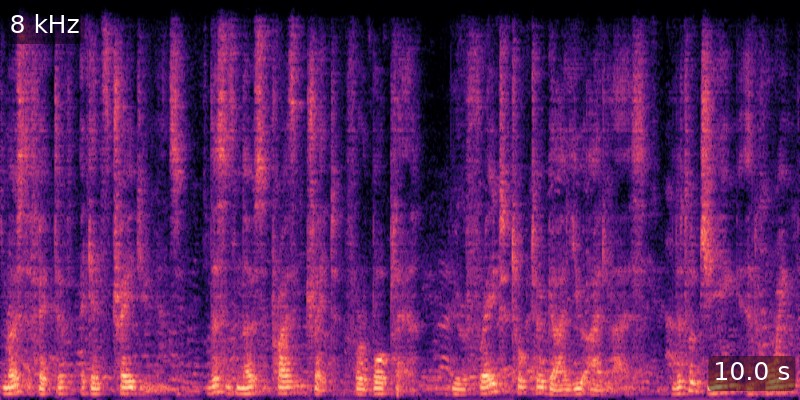}{WER 11.5\% (S2, D0, I1)}{most effective against trade unions this is no surprising trait for a \textcolor{errred}{board} \textcolor{errred}{player} youre afraid to talk to a guy you idolize little \textcolor{errred}{jeeing} and hawing}{cardborder}
    \hfill
    \spectrocard{TF-GridNet}{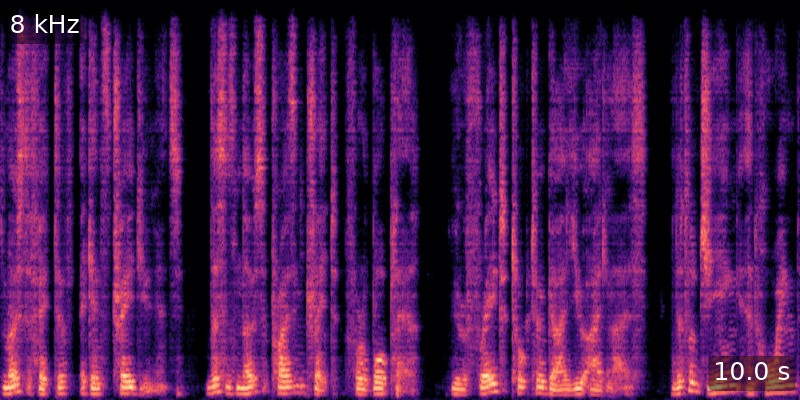}{WER 7.7\% (S1, D0, I1)}{most effective against trade unions this is no surprising trait for a \textcolor{errred}{board} \textcolor{errred}{player} youre afraid to talk to a guy you idolize little geeing and hawing}{cardborder}
    \hfill
    \spectrocard{PASE}{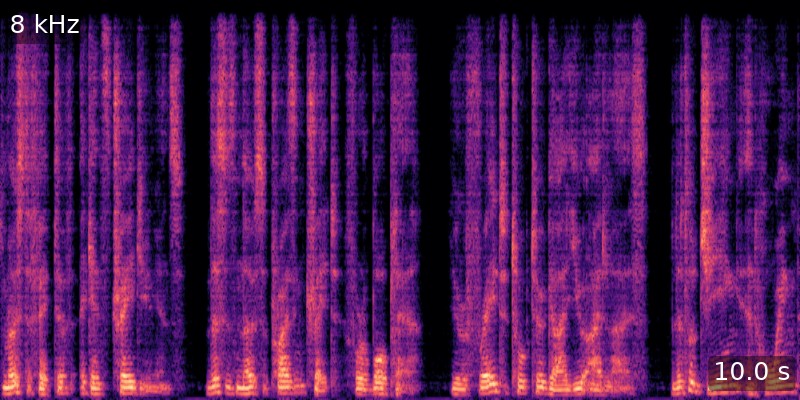}{WER 11.5\% (S2, D0, I1)}{most effective against trade unions this is no surprising trait for a \textcolor{errred}{board} \textcolor{errred}{player} youre afraid to talk to a guy you idolize little \textcolor{errred}{ging} and hawing}{cardborder}\par\smallskip
  \noindent
    \spectrocard{StuPASE}{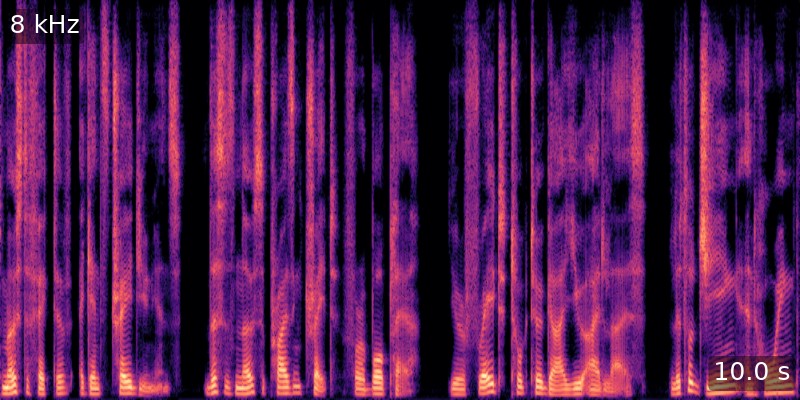}{WER 0.0\% (S0, D0, I0)}{most effective against trade unions this is no surprising trait for a ballplayer youre afraid to talk to a guy you idolize little geeing and hawing}{cardborder}
    \hfill
    \spectrocard{UniPASE}{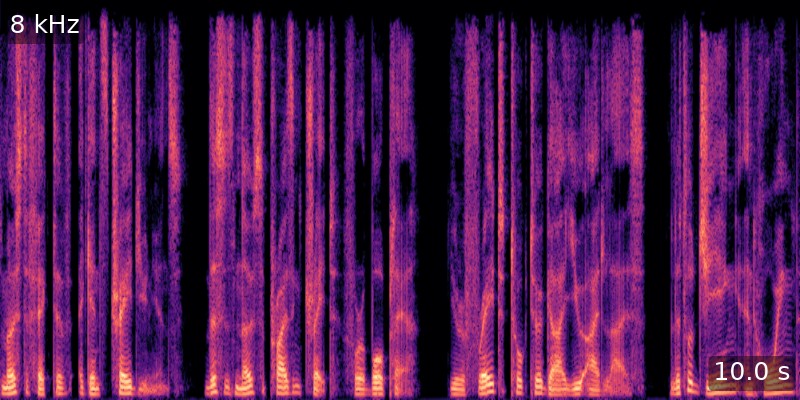}{WER 0.0\% (S0, D0, I0)}{most effective against trade unions this is no surprising trait for a ballplayer youre afraid to talk to a guy you idolize little geeing and hawing}{cardborder}
    \hfill
    \spectrocard{SenSE}{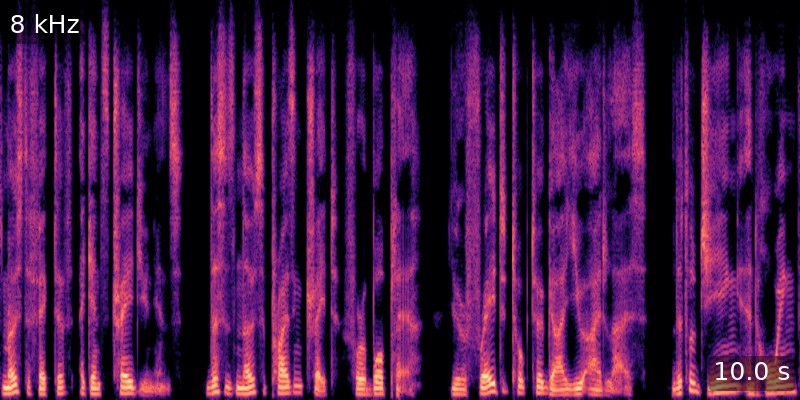}{WER 0.0\% (S0, D0, I0)}{most effective against trade unions this is no surprising trait for a ballplayer youre afraid to talk to a guy you idolize little geeing and hawing}{cardborder}\par\smallskip
  \noindent
    \spectrocard{UniSE}{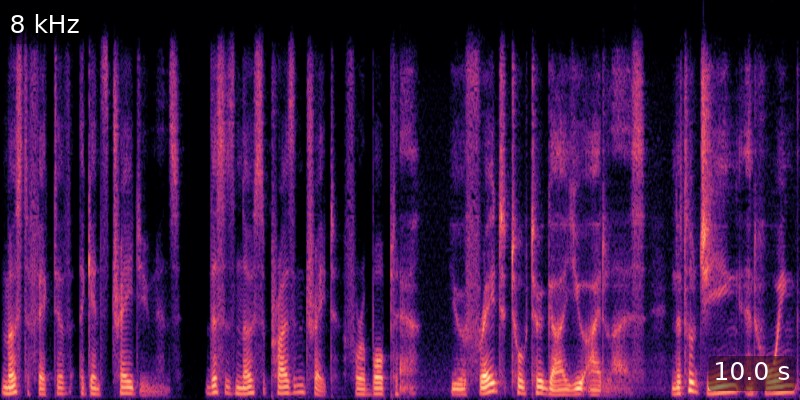}{WER 15.4\% (S3, D1, I0)}{most effective against \textcolor{errred}{tradeunions} this is no surprising trait for a ballplayer youre afraid to talk to a guy you idolize little \textcolor{errred}{jeeing} and \textcolor{errred}{hoeing}}{cardborder}
    \hfill
    \spectrocard{LIFT-SE (Ours)}{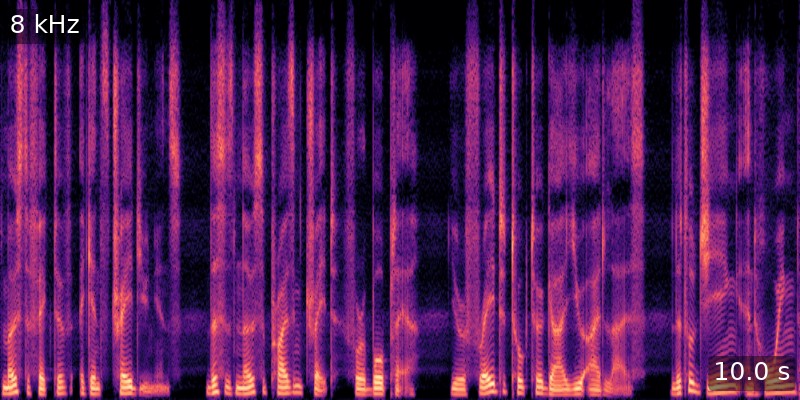}{WER 0.0\% (S0, D0, I0)}{most effective against trade unions this is no surprising trait for a ballplayer youre afraid to talk to a guy you idolize little geeing and hawing}{oursborder}
    \hfill
    \spectrocard{Clean}{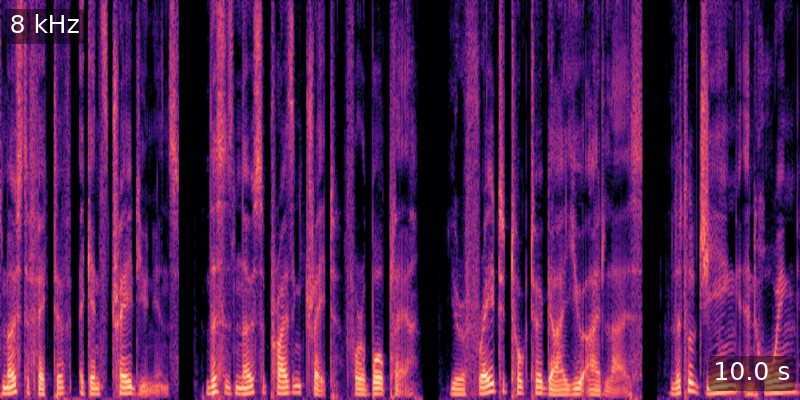}{Reference transcript}{most effective against trade unions this is no surprising trait for a ballplayer youre afraid to talk to a guy you idolize little geeing and hawing}{cardborder}
  \caption{Spectrograms and Whisper large-v3 transcripts of DNS1 \textit{with-reverb} utterance \texttt{fileid\_21}.}
  \label{fig:cards_3}
\end{figure*}

\begin{figure*}[!ht]
  \centering
  \footnotesize
  \noindent
    \spectrocard{Noisy}{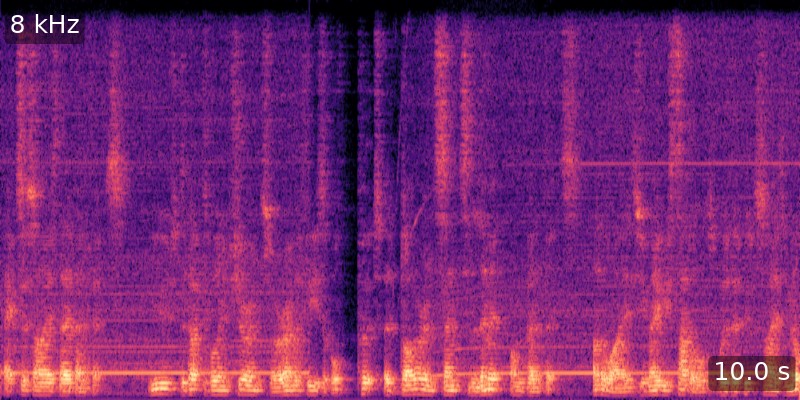}{WER 12.0\% (S2, D0, I1)}{exercise their benefits use deductible insurance wherever feasible put a dollar and \textcolor{errred}{cent} limit on benefits otherwise you may be saddled with a \textcolor{errred}{bootsized} milk \textcolor{errred}{cart}}{cardborder}
    \hfill
    \spectrocard{TF-GridNet}{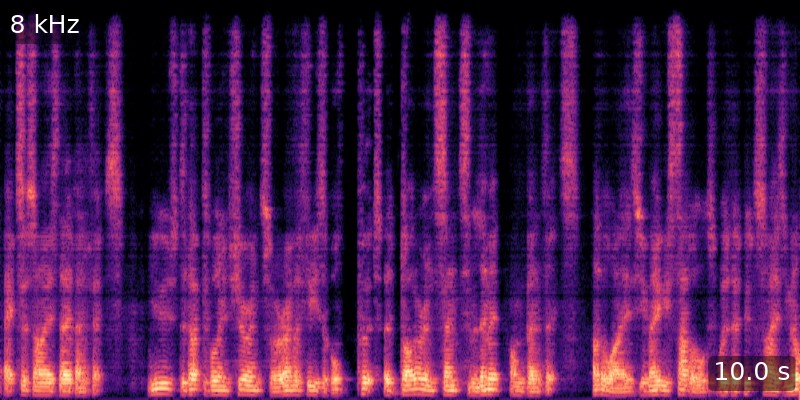}{WER 8.0\% (S2, D0, I0)}{exercise their benefits use deductible insurance wherever feasible put a dollar and \textcolor{errred}{cent} limit on benefits otherwise you may be saddled with a goodsized \textcolor{errred}{mil}}{cardborder}
    \hfill
    \spectrocard{PASE}{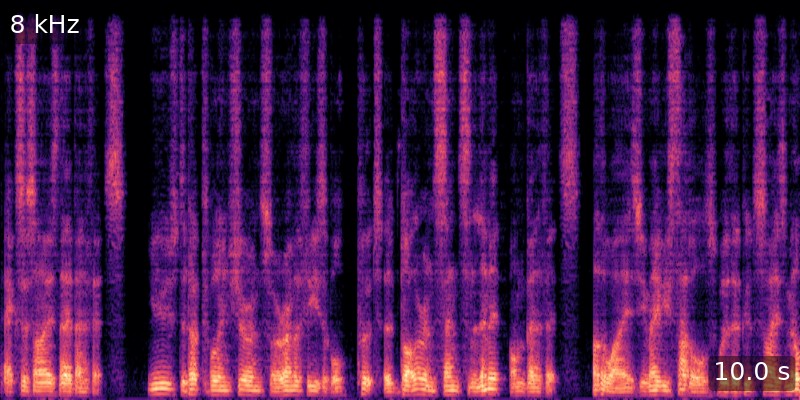}{WER 12.0\% (S1, D0, I2)}{exercise their benefits use deductible insurance \textcolor{errred}{or} \textcolor{errred}{are} \textcolor{errred}{more} feasible put a dollar and cents limit on benefits otherwise you may be saddled with a goodsized milk}{cardborder}\par\smallskip
  \noindent
    \spectrocard{StuPASE}{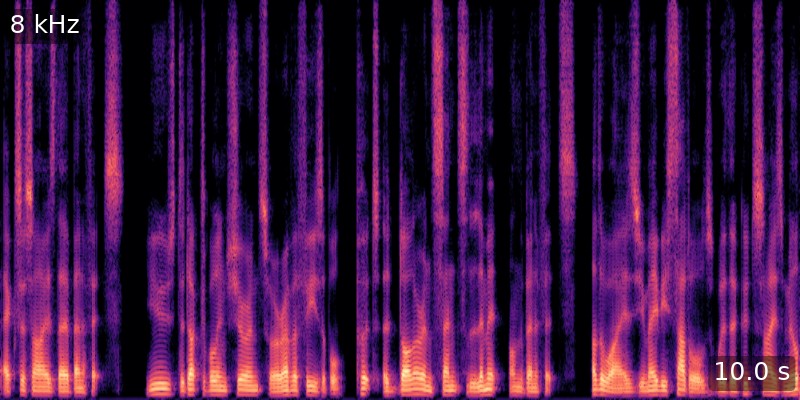}{WER 12.0\% (S1, D0, I2)}{exercise their benefits use deductible insurance wherever feasible put a dollar and cents limit on \textcolor{errred}{the} benefits otherwise you may be saddled with a \textcolor{errred}{good} \textcolor{errred}{sized} milk}{cardborder}
    \hfill
    \spectrocard{UniPASE}{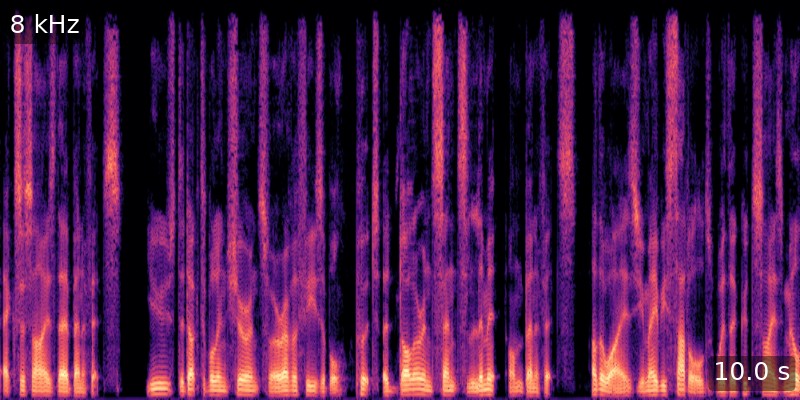}{WER 0.0\% (S0, D0, I0)}{exercise their benefits use deductible insurance wherever feasible put a dollar and cents limit on benefits otherwise you may be saddled with a goodsized milk}{cardborder}
    \hfill
    \spectrocard{SenSE}{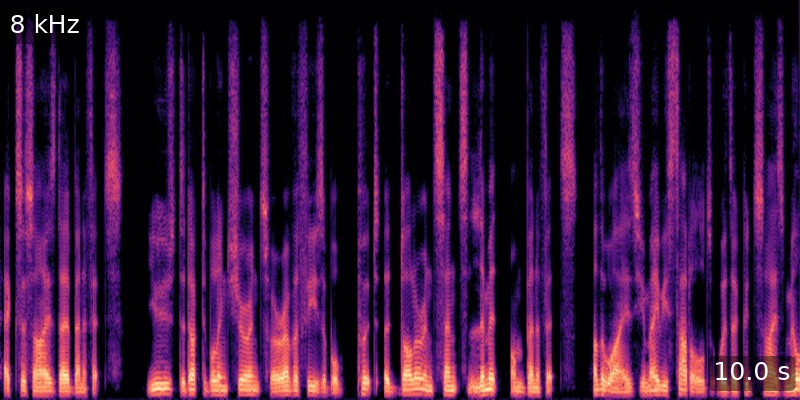}{WER 4.0\% (S0, D1, I0)}{exercise their benefits use deductible insurance wherever feasible put a dollar and cents limit on benefits otherwise you may be saddled with a goodsized}{cardborder}\par\smallskip
  \noindent
    \spectrocard{UniSE}{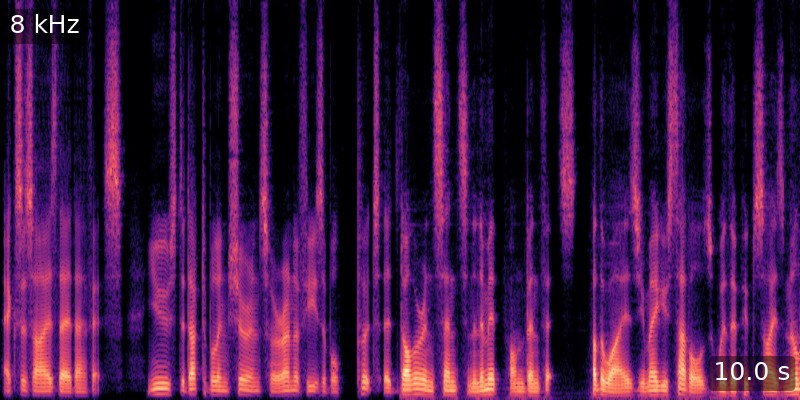}{WER 36.0\% (S4, D1, I4)}{exercise their benefits use deductible insurance \textcolor{errred}{for} \textcolor{errred}{a} \textcolor{errred}{runner} feasible put a dollar and cents \textcolor{errred}{in} \textcolor{errred}{the} limit on benefits otherwise you may \textcolor{errred}{use} \textcolor{errred}{saddles} with a \textcolor{errred}{poopsized}}{cardborder}
    \hfill
    \spectrocard{LIFT-SE (Ours)}{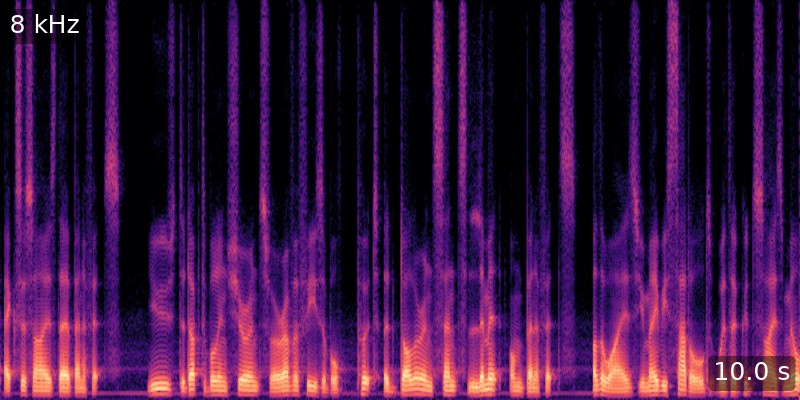}{WER 8.0\% (S1, D0, I1)}{exercise their benefits use deductible insurance wherever feasible put a dollar and cents limit on benefits otherwise you may be saddled with a \textcolor{errred}{good} \textcolor{errred}{size} milk}{oursborder}
    \hfill
    \spectrocard{Clean}{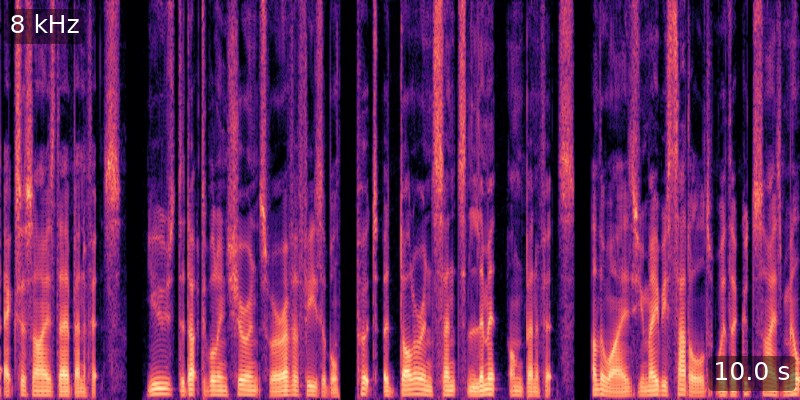}{Reference transcript}{exercise their benefits use deductible insurance wherever feasible put a dollar and cents limit on benefits otherwise you may be saddled with a goodsized milk}{cardborder}
  \caption{Spectrograms and Whisper large-v3 transcripts of DNS1 \textit{with-reverb} utterance \texttt{fileid\_73}.}
  \label{fig:cards_4}
\end{figure*}

\Crefrange{fig:cards_1}{fig:cards_4} show four utterances from the DNS1 \textit{with-reverb} test set, one per figure. Each card contains the spectrogram of one system output (0--8~kHz, normalized to its own peak with an $80$~dB range), the WER of its Whisper large-v3 transcript with the numbers of substituted (S), deleted (D), and inserted (I) words, and the transcript itself. The test set provides no transcripts, so the reference is the Whisper transcript of the clean (anechoic) recording, as in the dWER of \cref{sec:metrics}. Transcripts are shown after the text normalization used for scoring (lower-casing and removal of punctuation), and words that are substituted or inserted relative to the reference are marked in red.

\end{document}